\documentclass[prx,
nofootinbib,showpacs,amsmath,amssymb,superscriptaddress]{revtex4-2}

\usepackage{physics,amsmath,amsfonts,amssymb,graphicx,float,comment}
\usepackage{mathtools}
\usepackage[dvipsnames]{xcolor}
\usepackage{dsfont}

\def\>{\rangle}
\def\<{\langle}

\newcommand{\bbkk}[2]{\<\!\<{#1}|{#2}\>\!\>}

\newcommand{\kket}[1]{|{#1}\>\!\>}
\newcommand{\bbra}[1]{\<\!\<{#1}|}

\newcommand{\bes} {\begin{subequations}}
\newcommand{\ees} {\end{subequations}}
\newcommand{\beq}{\begin{equation}}
\newcommand{\eeq}{\end{equation}}

\newtheorem{theorem}{Theorem}[section]

\newtheorem{definition}{Definition}[section]

\usepackage{hyperref}
\hypersetup{
    colorlinks=true,
    linkcolor=blue,
    citecolor=red,      
    urlcolor=cyan,
}

\usepackage[capitalise]{cleveref}

\begin{document}
\title{Compressed Sensing Shadow Tomography} 
\author{Joseph Barreto}
\affiliation{Department of Physics and Astronomy, University of Southern California, Los Angeles, California 90089, USA}
\affiliation{Center for Quantum Information Science \& Technology, University of Southern California, Los Angeles, California 90089, USA}
\author{Daniel Lidar}
\affiliation{Department of Physics and Astronomy, University of Southern California, Los Angeles, California 90089, USA}
\affiliation{Center for Quantum Information Science \& Technology, University of Southern California, Los Angeles, California 90089, USA}
\affiliation{Department of Electrical and Computer Engineering, University of Southern California, Los Angeles, California 90089, USA}
\affiliation{Department of Chemistry, University of Southern California, Los Angeles, California 90089, USA}
\affiliation{Quantum Elements, Inc., Thousand Oaks, California, USA}
\date{\today}

\begin{abstract}
Estimating many local expectation values over time is a central measurement bottleneck in quantum simulation and device characterization. We study the task of reconstructing the Pauli-signal matrix
$S_{ij}=\Tr \big(O_i \rho(t_j)\big)$
for a collection of $M$ low-weight Pauli observables $\{O_i\}_{i=1}^M$ over $N$ timesteps $\{t_j\}_{j=1}^N$, while minimizing the total number of device shots. We propose a \emph{Compressed Sensing Shadow Tomography} (CSST) protocol that combines two complementary reductions. First, local classical shadows reduce the observable dimension by enabling many Pauli expectation values to be estimated from the same randomized snapshots at a fixed time. Second, compressed sensing reduces the time dimension by exploiting the fact that many expectation-value traces are spectrally sparse or compressible in a unitary (e.g., Fourier) transform basis.  Operationally, CSST samples $m\ll N$ timesteps uniformly at random, collects shadows only at those times, and then reconstructs each length-$N$ signal via standard $\ell_1$-based recovery in the unitary transform domain.  We provide end-to-end guarantees that explicitly combine shadow estimation error with compressed sensing recovery bounds.  For exactly $s$-sparse signals in a unitary transform basis, we show that $m=\mathcal{O} \left(s\log^2 s \log N\right)$ random timesteps suffice (with high probability), leading to total-shot savings scaling as $\widetilde{\Theta}(N/s)$ (i.e., up to polylogarithmic factors) relative to collecting shadows at all $N$ timesteps.  For approximately sparse signals, the reconstruction error decomposes into a compressibility (tail) term plus a noise term. We present numerical experiments on noisy many-qubit dynamics that support strong Fourier compressibility of Pauli traces and demonstrate substantial shot reductions with accurate reconstruction.

\end{abstract}

\maketitle
\tableofcontents

\section{Introduction}
\label{sec:intro}

Understanding the behavior of many-body quantum systems over time is relevant to theoreticians studying information flow and scrambling, to experimentalists characterizing engineered Hamiltonians and noise processes in quantum hardware, and to practitioners simulating dynamics in physical systems such as atoms and molecules.  
In all of these settings, the bottleneck is often not state preparation or time evolution, but \emph{measurement}: acquiring sufficiently many samples to reliably estimate a large collection of observables across many timesteps.

In this work we study the following measurement task.  Let $\rho(t)$ denote the state of an $n$-qubit system at time $t$ in a Hilbert space $\mathcal{H} \cong \mathbb{C}^{d}$ of dimension $d=2^n$, and fix a set of $M$ Hermitian Pauli observables $\{O_i\}_{i=1}^M$ together with a set of $N$ timesteps $\{t_j\}_{j=1}^N$.  Our goal is to estimate the data matrix $S\in\mathbb{R}^{M\times N}$ with entries $S_{ij} \equiv \Tr(O_i \rho(t_j))$.
We consider the open system setting, where the state obeys a master equation $\dot{\rho} = \mathcal{L}(\rho)$ with Lindblad generator $\mathcal{L}$~\cite{breuerTheoryOpenQuantum2009}, treating the device as a black box that can be repeatedly run to prepare $\rho(t_j)$ and return measurement samples at the designated timestep $t_j$.
We  focus on the common and practically important case where the $O_i$ are low-weight Pauli strings (i.e., acting nontrivially on a small number of qubits),
since such observables underlie many real-time spectral evolution schemes~\cite{shenEstimatingEigenenergiesQuantum2025, chanAlgorithmicShadowSpectroscopy2025} and hardware characterization routines \cite{samachLindbladTomographySuperconducting2022, baireyLearningDynamicsOpen2020, onoratiFittingQuantumNoise2023}.  When $M$ is large, one can sometimes reduce measurement overhead by grouping commuting (or qubit-wise commuting) Pauli strings~\cite{reggioFastPartitioningPauli2024, verteletskyiMeasurementOptimizationVariational2020}, or by using randomized measurement protocols that support many observables simultaneously~\cite{elbenRandomizedMeasurementToolbox2022}.

A particularly flexible randomized approach is classical shadow tomography (ST) \cite{huangPredictingManyProperties2020}.  At a \emph{single} timestep $t_j$, local classical shadows provide a ``measure first, ask later'' guarantee: with a number of randomized snapshots\footnote{We use ``snapshot'' and ``shot'' interchangeably; each snapshot corresponds to one randomized product measurement on all 
$n$ qubits.} that scales only logarithmically in the number of target observables (and exponentially only in the maximum Pauli weight), one can estimate all $M$ values $\{\Tr(O_i\rho(t_j))\}_{i=1}^M$ to a prescribed accuracy with high probability.  However, when the object of interest is a \emph{time series}, the naive use of ST still requires repeating this procedure at every timestep.  The total measurement cost therefore grows at least linearly with $N$, which quickly becomes prohibitive when high temporal resolution is desired.

The central observation motivating this work is that observable time series are often \emph{structured}.  In many closed-system and open-system settings, expectation values $\Tr \left(O_i \rho(t)\right)$
are well-approximated by superpositions of relatively few oscillatory (and possibly weakly decaying) modes.  As a result, the discretized signals $\{S_{ij}\}_{j=1}^N$ are frequently sparse or compressible in a discrete Fourier or cosine transform (DFT or DCT) domain.  This spectral simplicity suggests a different measurement strategy: instead of spending shots uniformly across all $N$ timesteps, one can measure at only a \emph{random subset} of times and reconstruct the full-length signals using compressed sensing~\cite{candesRobustUncertaintyPrinciples2004, candesStableSignalRecovery2005, candesOptimalSignalRecovery2006, foucartMathematicalIntroductionCompressive2013}.

\begin{figure}
    \centering
    \includegraphics[width=0.9\linewidth]{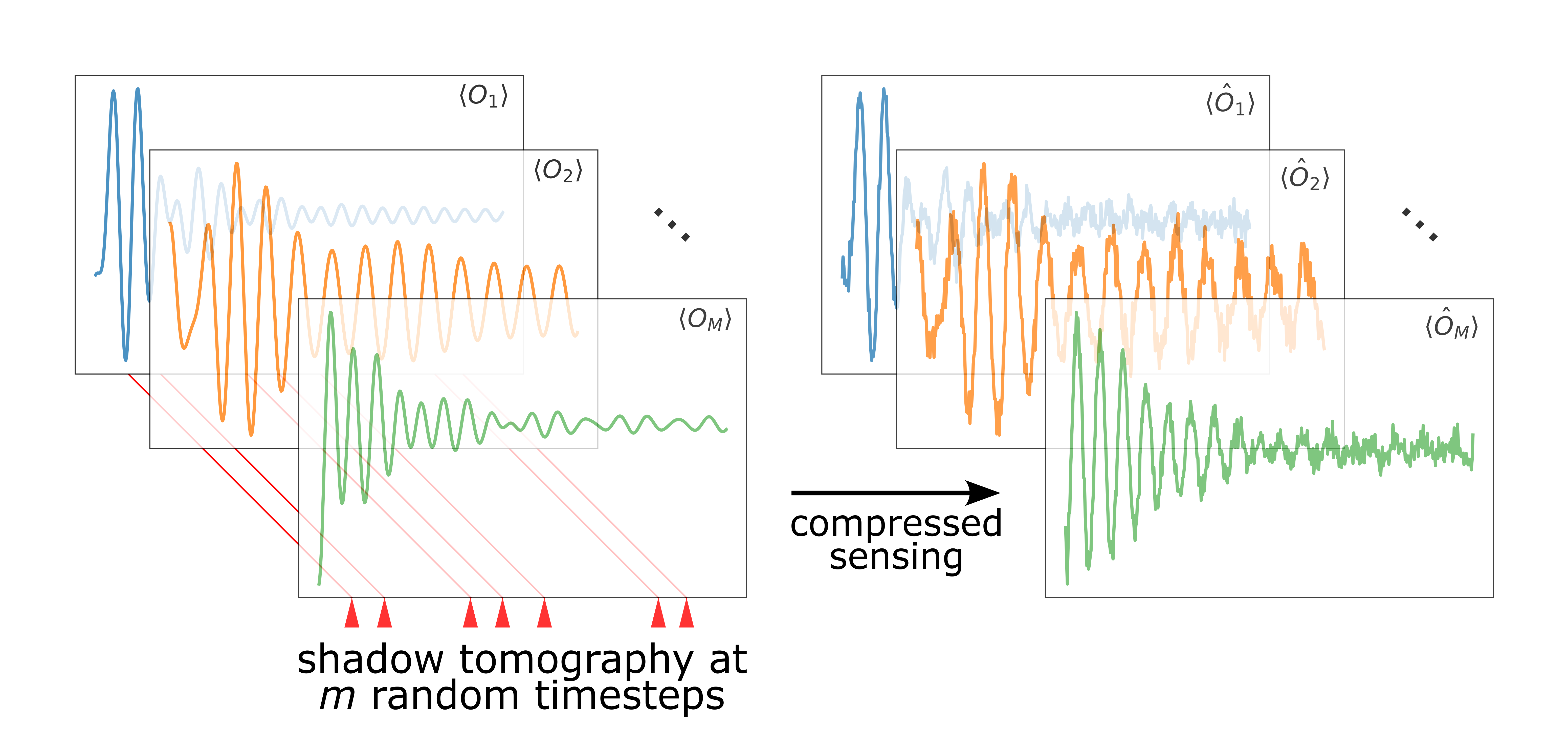}
    \caption{Schematic of the CSST data-processing pipeline. We sample $m\ll N$ timesteps uniformly at random (red markers) and, at each sampled time, collect local classical-shadow snapshots that enable simultaneous estimation of the $M$ target low-weight Pauli expectation values $\{\Tr(O_i\rho(t_j))\}_{i=1}^M$. For each observable $O_i$, the resulting time-subsampled signal is then reconstructed on the full length-$N$ grid using Fourier/DCT-domain compressed sensing.}
    \label{fig:csst}
\end{figure}

Motivated by this spectral compressibility, we propose a Compressed Sensing Shadow Tomography (CSST) protocol that measures local classical shadows at $m\ll N$ randomly chosen timesteps and reconstructs each length-$N$ signal via Fourier/cosine-domain compressed sensing; see \cref{fig:csst}.
We provide end-to-end guarantees that combine (a) finite-sample shadow tomography error bounds at the sampled timesteps with (b) compressed sensing recovery guarantees for time-subsampled signals.  Our bounds quantify how the required number of sampled timesteps $m$, the per-timestep snapshot count, the maximum Pauli weight, and the target failure probability must scale to ensure accurate reconstruction \emph{uniformly over} a family of $M$ observables. We compare against a baseline that runs local classical shadows independently at all $N$ timesteps (with the same target accuracy and failure probability), allocating snapshots uniformly across time. Our theory makes precise when time-subsampling yields a net advantage over this baseline, and shows that this advantage scales with the ratio between $N$ and the effective sparsity/compressibility of the signals.

We first analyze an \emph{idealized exactly sparse benchmark} to make the best-case scaling transparent. 
Assume each length-$N$ observable signal 
\beq
    \mathbf{s}_i = (\Tr[O_i \rho(t_1)],\dots,\Tr[O_i \rho(t_N)])^T \in\mathbb{R}^N
\eeq
is $s$-sparse in a unitary transform domain $F$ (e.g., DFT/DCT), i.e., $F\mathbf{s}_i$ has at most $s$ nonzero entries. Let $\mathbf{s}_i^*$ denote the reconstructed signal, and fix a target reconstruction error $\mathrm{RMSE}(\mathbf{s}_i,\mathbf{s}_i^*) \le \epsilon_{\mathrm{RMS}}$ (i.e., $\|\mathbf{s}_i-\mathbf{s}_i^*\|_2 \le \sqrt{N} \epsilon_{\mathrm{RMS}}$)
and failure probability at most $\delta$ (over both the random timestep subsampling and the random shadow measurement settings/outcomes), uniformly over a family of $M$ Pauli observables $O_i$ of weight at most $w$. Let $N_{\mathrm{tot}}^{\mathrm{ST}}$ denote the total number of measurement shots under the baseline ST protocol, i.e., at all $N$ timesteps. Instead, our protocol uses $m = \mathcal{O}\left(s\log^2 s  \log N\right)$ random timesteps to reconstruct the full length-$N$ signal. This protocol requires $N_{\mathrm{tot}}^{\mathrm{CS}}$ total measurement shots. With this choice for $m$, we demonstrate (see \cref{sec:exact-sparse}) that the shot ratio satisfies
\begin{align}
    \label{eq:shot-ratio}
    \frac{N_{\mathrm{tot}}^{\mathrm{ST}}}{N_{\mathrm{tot}}^{\mathrm{CS}}} = \widetilde{\Theta} \left(\frac{N}{s}\right),
\end{align}
where the tilde denotes scaling up to polylogarithmic factors.
When the effective sparsity $s$ is much smaller than the time-series length $N$, subsampling in time and reconstructing via compressed sensing reduces the total shot cost by a factor that grows essentially linearly with $N/s$.

In realistic open-system dynamics, however, exact sparsity in a fixed DFT/DCT grid is uncommon; the practically
relevant regime is \emph{approximate} sparsity/compressibility, for which we find similar improvements, though with caveats determined by the degree of compressibility.

These results can conceptually be understood as a combination of two phenomena.  First, signals which are sparse in some sense should require fewer measurements to estimate altogether, and compressed sensing offers favorable bounds in terms of this sparsity level.  Second, when measurement noise is not so large that it overwhelms the signal’s sparse/compressible structure in the transform domain, sparse recovery can effectively denoise by enforcing that structure. 

This paper is organized as follows.
\cref{sec:background} reviews related work in ST, compressed sensing, and in quantum spectral estimation from time-domain data. It also defines our open system model.
\cref{sec:csst_bounds} presents our protocol and the main end-to-end error/shot-complexity guarantees. We consider both the case of exact $s$-sparse signals and the case of approximate $s$-sparse signals.
\cref{sec:numerics} reports numerical experiments that corroborate the predicted savings and illustrate when time-subsampling is most beneficial. As a preview of the advantage achievable via CSST over ST, see \cref{fig:cs_example}.
We conclude in \cref{sec:conc} with an overview and future directions. Additional technical details in support of the main text are presented in the Appendix, which includes a complete table of our notational conventions (see \cref{tab:notation}).

\section{Background}
\label{sec:background}

For each observable $i\in\{1,\dots,M\}$, let $\mathbf{s}_i\in\mathbb{R}^N$ denote the corresponding length-$N$
time series (the $i$'th row of $S$), i.e., $(\mathbf{s}_i)_j = S_{ij}$, and let $\widehat{\mathbf{s}}_i$ denote the corresponding estimated signal.
Throughout, we quantify time-series reconstruction accuracy using the root-mean-square (RMS) error over time,
\begin{equation}
    \mathrm{RMSE}(\mathbf{s}_i,\widehat{\mathbf{s}}_i) \equiv \frac{1}{\sqrt{N}}\|\mathbf{s}_i-\widehat{\mathbf{s}}_i\|_2.
    \label{eq:rmse_def}
\end{equation}
Equivalently, an RMSE target $\mathrm{RMSE}(\mathbf{s}_i,\widehat{\mathbf{s}}_i)\le \epsilon_{\mathrm{RMS}}$ is
the same as an $\ell_2$ target $\|\mathbf{s}_i-\widehat{\mathbf{s}}_i\|_2 \le \sqrt{N} \epsilon_{\mathrm{RMS}}$.
We adopt RMSE since it matches our numerical experiments and corresponds to an average per-timestep error.

\subsection{Prior Work}
\label{sec:prior}

\subsubsection{Shadow Tomography}
Classical shadow tomography (ST)~\cite{huangPredictingManyProperties2020} provides a randomized ``measure first, ask later'' data structure for estimating many observables of a quantum state from a single measurement dataset.  In our setting, at a fixed timestep $t_j$ we collect local classical-shadow snapshots of $\rho(t_j)$ and then use the resulting dataset to estimate a family of Pauli expectation values $\{S_{ij}\}_{i=1}^M$.

For a family of $M$ Pauli strings $\{O_i\}_{i=1}^M$ with maximum Pauli weight at most $w$, ST yields uniform (simultaneous) accuracy guarantees at a fixed timestep: with $N_{\text{ST}}$ snapshots of $\rho(t_j)$ one can construct estimators $\{\widehat{S}_{ij}\}_{i=1}^M$ such that, with probability at least $1-\delta$,
\begin{align}
    \max_{i\in\{1,\dots,M\}} \left|\widehat{S}_{ij}-\Tr \left(O_i\rho(t_j)\right)\right|\le \epsilon,
\end{align}
where the required snapshot count scales as\footnote{This scaling is also achieved via quantum overlapping tomography, which is closely related to ST specialized to low-weight Pauli strings~\cite{cotlerQuantumOverlappingTomography2020}.}

\begin{align}
    \label{eq:N_ST_0}
    N_{\text{ST}} = \mathcal{O}\left(\frac{3^w}{\epsilon^2}\log\frac{M}{\delta}\right).
\end{align}

Here the Pauli weight $w(O)$ is the number of non-identity single-qubit Paulis in the tensor product expansion
\begin{align}
    \label{eq:Pauli-string}
    O  =  \bigotimes_{k=1}^n P_{k},
    \quad
    P_k \in \{I,X,Y,Z\},
\end{align}
i.e.,
\begin{align}
    w(O) = \sum_{k=1}^n (1-\delta_{P_k, I}).
\end{align}
Conceptually, ST avoids reconstructing $\rho(t_j)$ and instead enables post-processing estimation of many Pauli expectation values (up to the chosen weight cutoff) from the same set of randomized local measurement settings.  We review the estimators and concentration bounds used in this work in \cref{app:shadow}.

A well-known caveat is that if the target observable family is highly structured (e.g., heavily biased toward certain Paulis), then one can often do substantially better than fully randomized measurements via derandomization or deterministic measurement design~\cite{huangEfficientEstimationPauli2021}.  Nevertheless, for large unstructured families and especially when one wants \emph{many} observables, ST provides a flexible baseline.  In the time-series setting, however, the naive use of ST repeats the above procedure at each desired timestep, making the overall shot cost scale at least linearly in the number of timesteps.  Our protocol addresses this by measuring only a random subset of times and exploiting spectral structure of the resulting signals via compressed sensing.

\subsubsection{Compressed Sensing in Quantum Tomography}

Compressed sensing has been widely used in quantum information to reduce measurement costs when the object of interest has exploitable structure, e.g., low-rank states or sparse parameterizations.  Notable examples include compressed-sensing state tomography~\cite{grossQuantumStateTomography2010}, process tomography~\cite{shabaniEfficientMeasurementQuantum2011}, Hamiltonian learning~\cite{shabaniEstimationManybodyQuantum2011, maLearning$k$bodyHamiltonians2024}, and Lindbladian tomography~\cite{dobryninCompressedsensingLindbladianQuantum2024}, including experimental demonstrations~\cite{riofrioExperimentalQuantumCompressed2017}.  In these works, one aims to reconstruct a large operator (a state, channel, or generator) from fewer measurements by assuming global structure.

For \emph{spectral} estimation, reconstructing the full generator is often undesirable even if it is information-theoretically possible: numerical stability can be poor because eigenvalue errors can be amplified by eigenvector conditioning.  Concretely, if one estimates the dynamical generator $L$ (\cref{eq:L}) and obtains $\hat{L}=L+E$ with $L = V\Lambda V^{-1}$, then the Bauer-Fike theorem~\cite{hornMatrixAnalysis2012} implies that for any $\hat{\lambda}\in\sigma(\hat{L})$ there exists $\lambda\in\sigma(L)$ such that
\begin{align}
    |\hat{\lambda}-\lambda|\le \kappa(V) \|E\|,
\end{align}
where $\kappa(V)=\|V\| \|V^{-1}\|$ for any induced matrix norm.  For generic open systems there is no reason for $\kappa(V)$ to be small, and it can become very large near exceptional points~\cite{mingantiQuantumExceptionalPoints2019}.  Thus, accurate eigenvalue inference via ``estimate-then-diagonalize'' may require extremely small estimation error $\|E\|$, which is the main reason for not attempting to just reconstruct the generator of the dynamics.

A common alternative in the closed-system setting is to project dynamics into a low-dimensional Krylov (or related) subspace and solve a reduced eigenvalue problem \cite{cortesQuantumKrylovSubspace2022, epperlyTheoryQuantumSubspace2022}.  While this avoids explicitly representing the full generator, it still requires estimating many overlap quantities and can inherit conditioning issues when the prepared basis becomes nearly linearly dependent.

In this work, we use compressed sensing in a different way: we do not apply it to operator reconstruction.  Instead, we apply it to the \emph{time axis}, leveraging sparsity/compressibility of expectation-value time series in a Fourier-/cosine-type basis to reduce the number of timesteps at which measurements are required, which is one of compressed sensing's original applications \cite{candesRobustUncertaintyPrinciples2004, candesStableSignalRecovery2005}.

\subsubsection{Real-Time Methods}

Rather than reconstructing the full generator, a broad class of ``real-time'' spectral estimation methods extract spectral information directly from time-domain data.  Examples include methods based on time-series fitting and subspace identification (e.g., matrix pencil, ESPRIT, and dynamic mode decomposition) applied to quantum signals~\cite{sommaQuantumEigenvalueEstimation2019, obrienQuantumPhaseEstimation2019, hangleiterRobustlyLearningHamiltonian2024, shenEfficientMeasurementDrivenEigenenergy2024, shenEstimatingEigenenergiesQuantum2025, kanekoForecastingLongtimeDynamics2025, chanAlgorithmicShadowSpectroscopy2025}.  These approaches effectively fit measured signals to a model that is a linear combination of (possibly damped) complex exponentials (matching the modal expansion derived in \cref{eq:obs_exp}, below).

In these approaches, the classical post-processing is typically not the bottleneck; the dominant cost is the number of measurement shots needed to estimate the signal at sufficiently many time points with adequate SNR.  Our contribution is to reduce this \emph{measurement} cost while leaving the downstream signal-processing task unchanged: we (i) use local classical shadows to estimate many Pauli expectation values at selected times, and (ii) exploit Fourier/DCT sparsity or compressibility to reconstruct full time series from a random time subsample via compressed sensing.

Real-time approaches differ in the precise signal that is measured (e.g., interference-based autocorrelations versus expectation values).  Since our protocol operates on expectation values of low-weight Pauli strings, it applies directly to both unitary and Lindbladian dynamics using only local measurements, and it is compatible with open-system settings where oscillation frequencies and decay rates are governed by the Liouvillian spectrum.

\subsection{Signal model and Fourier/DCT compressibility}

In the Markovian setting of interest to us here, the state evolves according to the Lindblad equation in the Schr\"{o}dinger picture:
\begin{align}
    \dot{\rho} = \mathcal{L}(\rho) \implies
    \rho(t) = (\exp(t\mathcal{L}))(\rho_0) \implies \kket{\rho(t)} = \exp(tL) \kket{\rho_0}
\end{align}
where $\rho \rightarrow \kket{\rho}$ denotes the vectorization of an operator to a vector, and $\mathcal{L} \rightarrow L$ of a superoperator to a matrix. The explicit form of the vectorized Lindbladian is found using Roth's lemma~\cite{rothDirectProductMatrices1934} \begin{align}
\label{eq:L}
    L &= -i(I \otimes H - H^T \otimes I) + \sum_{i} \gamma_{i} \Big(  L^*_{i} \otimes L_{i} - \frac{1}{2} I \otimes L^{\dagger}_{i} L_{i} - \frac{1}{2} (L_{i}^{\dagger} L_{i})^T \otimes I\Big) ,
\end{align}
where $H$ is the Hamiltonian and the $L_i$ are the Lindblad operators.

Equivalently, observables evolve under the adjoint generator $L^\dagger$. 
In this notation, the Hilbert-Schmidt inner product of two arbitrary operators $A,B$ becomes
\begin{align}
    \< A,B \>_{\text{HS}} = \Tr(A^\dagger B) = \bbkk{A}{B}
\end{align}
Assuming that $L$ is diagonalizable (i.e., no Jordan blocks) \cite{endeAlmostAllQuantum2024}, we can write $L = P \Lambda P^{-1}$, where $P = \sum_{k=0}^{d^2 - 1} \kket{r_k}\bbra{k}$ and $P^{-1} = \sum_{k=0}^{d^2 - 1} \kket{k}\bbra{l_k}$.\footnote{In the presence of Jordan blocks, signals acquire $\text{poly}(t)$ prefactors multiplying exponentials, which generally increases effective bandwidth and can reduce sparsity; our numerics (see \cref{sec:numerics}) suggest compressibility often still holds in practice.} Here, $k$ is an index labeling the standard basis, and $\{\kket{r_k}\},\{\bbra{l_k}\}$ are the biorthonormal sets of right and left eigenvectors of the $d^2\times d^2$ Liouvillian (or Lindbladian) matrix $L$, i.e., $L\kket{r_k}=\lambda_k\kket{r_k}$ and $\bbra{l_k}L=\lambda_k\bbra{l_k}$, where $\lambda_k \in \mathbb{C}$ and $\Re[\lambda_k] \le 0$.   

Define the (scalar) signal
\begin{align}
    s_O(t) \equiv \Tr(O\rho(t)) = \bbkk{O}{\rho(t)}
    = \bbra{O} e^{tL}\kket{\rho_0} .
\end{align}
Expanding $\rho_0$ and $O$ into $L$'s right and left eigenvectors, we have
\begin{align}
    \kket{\rho_0} = \sum_n b_n \kket{r_n} \quad \bbra{O} = \sum_m a_{m} \bbra{l_m}
\end{align}
Using the bi-orthogonality property $\< \< l_m \kket{r_n} = \delta_{mn}$, we obtain
\begin{align}
 \label{eq:obs_exp}
   s_O(t)
    = \sum_{k=0}^{d^2-1} e^{\lambda_k t} \bbkk{O}{r_k} \bbkk{l_k}{\rho_0}
      = \sum_{k=0}^{d^2 - 1} e^{\lambda_k t} a_k b_{k},
\end{align}

In the purely unitary case, $L$ reduces to the commutator generator and its eigenvalues are (imaginary) Bohr frequencies; weak dissipation perturbs these frequencies and introduces decay via negative real parts.  Crucially, although $L$ is $d^2\times d^2$, in many settings a given pair $(\rho_0,O)$ couples appreciably to only a subset of modes over the time window of interest, so the resulting signal is often well-approximated by a superposition of relatively few (weakly damped) oscillatory components.  This is the structural assumption we exploit: after discretization, many such signals are sparse or compressible in a Fourier-/cosine-type basis, making them amenable to reconstruction from randomly time-subsampled data via compressed sensing~\cite{candesRobustUncertaintyPrinciples2004, candesStableSignalRecovery2005, candesOptimalSignalRecovery2006, foucartMathematicalIntroductionCompressive2013}.

\section{Estimation via Shadow Tomography and Compressed Sensing}
\label{sec:csst_bounds}

\subsection{Method overview}

Our CSST protocol is as follows, as illustrated schematically in \cref{fig:csst}:
\begin{enumerate}
    \item Choose an index set $\Omega\subset\{1,\dots,N\}$ of size $m\ll N$ uniformly at random (without replacement), corresponding to sampled timesteps $\{t_j:j\in\Omega\}$.
    \item For each sampled timestep $t_j$ with $j\in\Omega$, collect $N_{\text{ST}}$ local classical-shadow snapshots of $\rho(t_j)$ (i.e., randomized product measurement settings), yielding a shadow dataset usable to estimate many low-weight Pauli expectation values.
    \item For each target Pauli string $O_i$, form the (noisy) time-subsampled measurements of its signal on $\Omega$ and reconstruct the full length-$N$ signal using compressed sensing in a DFT/DCT-type basis.
\end{enumerate}

In this section, we analyze the CSST protocol by combining standard guarantees from (i) local classical shadows and (ii) compressed sensing to obtain end-to-end bounds for estimating each observable time-series
\begin{align}
    \mathbf{s}_i \equiv (S_{i1},\dots,S_{iN})^T,\quad S_{ij} = \Tr \big(O_i \rho(t_j)\big),
\end{align}
where $\{O_i\}_{i=1}^M$ are Pauli strings of weight at most $w$ and $\{t_j\}_{j=1}^N$ are the timesteps of interest.
We quantify reconstruction accuracy using RMSE [\cref{eq:rmse_def}],
and our goal is to achieve $\mathrm{RMSE}(\mathbf{s}_i,\widehat{\mathbf{s}}_i)\le \epsilon_{\mathrm{RMS}}$ for all $i\in\{1,\dots,M\}$
with probability at least $1-\delta$.
(Equivalently, $\|\mathbf{s}_i-\widehat{\mathbf{s}}_i\|_2 \le \sqrt{N} \epsilon_{\mathrm{RMS}}$.)

We compare the CSST protocol to a baseline ``ST-at-all-times'' protocol for obtaining the estimates $\widehat{\mathbf{s}}_i$: perform ST independently at all $N$ timesteps and set $\widehat{\mathbf{s}}_i=(\widehat{S}_{i1},\dots,\widehat{S}_{iN})^T$.  The main result we obtain is a shot-count comparison between the two protocols (up to logarithmic factors), showing that \emph{CSST can significantly reduce measurement cost when the signals are sparse/compressible in the chosen transform domain}.

\subsection{Local classical shadows at a single timestep}

We use the following result:
\begin{theorem}[Local Classical Shadows \cite{huangPredictingManyProperties2020}]
    \label{th:st}
    Fix a state $\rho$ on $n$ qubits and a set of $M$ Pauli observables $\{O_i\}_{i=1}^M$ each with weight at most $w$. There exist estimators $\{\hat{o}_i\}_{i=1}^M$ constructed from $N_{\mathrm{ST}}$ local classical-shadow snapshots such that, with probability at least $1-\delta$,
    \begin{align}
    \max_{i\in\{1,\dots,M\}} \big|\Tr(O_i\rho)-\hat{o}_i\big| \le \epsilon_{\mathrm{ST}},
    \end{align}
    provided
    \begin{align}
    \label{eq:N_ST}
    N_{\mathrm{ST}} = \mathcal{O}\left(\frac{3^w}{\epsilon_{\mathrm{ST}}^2}\log\frac{M}{\delta}\right).
    \end{align}
\end{theorem}

\subsection{Baseline protocol: shadows at all $N$ timesteps}
Assume we run \cref{th:st} independently at each timestep $t_j$ to estimate all $M$ observables. Let $\hat{S}$ denote the resulting estimated dataset matrix and define the error matrix $E$ via
\begin{align}
    \hat{S} = S + E,\quad \hat{S}_{ij} = S_{ij} + e_{ij}.
\end{align}
For each fixed timestep $t_j$, define the event $A_j \equiv \left\{\max_{i\in\{1,\dots,M\}} |e_{ij}| \le \epsilon_{\mathrm{ST}}\right\}$.
By \cref{th:st}, we have $\Pr(A_j) \ge 1-\delta_t$, where $\delta_t$ is the per-timestep failure probability.
Equivalently $\Pr(A_j^c)\le \delta_t$ for the complement event $A_j^c$. We would like the bound to hold for all $N$ timesteps simultaneously, i.e.,
$A \equiv \bigcap_{j=1}^N A_j
= \left\{\max_{i,j} |e_{ij}| \le \epsilon_{\mathrm{ST}}\right\}$. The complement event is
$A^c = \bigcup_{j=1}^N A_j^c$,
so by the union bound,
\begin{align}
    \label{eq:baseline_union_bound}
    \Pr(A^c) = \Pr \left(\bigcup_{j=1}^N A_j^c\right)
    \le \sum_{j=1}^N \Pr(A_j^c)
    \le \sum_{j=1}^N \delta_t
    = N\delta_t.
\end{align}
Therefore, $\Pr(A) = 1-\Pr(A^c) \ge 1-N\delta_t$. Setting $\delta_t=\delta/N$ yields $\Pr(A)\ge 1-\delta$,
i.e., with probability at least $1-\delta$,
\begin{align}
    \max_{i,j} |e_{ij}| \le \epsilon_{\mathrm{ST}}.
\end{align}
Consequently, for each fixed observable $i$, define the length-$N$ error vector $\mathbf{e}_i^{\rm full}\equiv (e_{i1},\dots,e_{iN})^T$ and note that
\begin{align}
    \label{eq:e_i-ell2}
    \|\mathbf{e}_i^{\rm full}\|_2 = \left(\sum_{j=1}^N |e_{ij}|^2\right)^{1/2} \le \sqrt{N} \epsilon_{\mathrm{ST}}.
\end{align}
Dividing \cref{eq:e_i-ell2} by $\sqrt{N}$ yields an RMSE guarantee for each signal:
\begin{align}
    \mathrm{RMSE}(\mathbf{s}_i,\widehat{\mathbf{s}}_i)
    =
    \frac{1}{\sqrt{N}}\|\mathbf{e}_i^{\rm full}\|_2
    \le
    \epsilon_{\mathrm{ST}}.
\end{align}
Therefore, under the baseline protocol, to guarantee $\mathrm{RMSE}(\mathbf{s}_i,\widehat{\mathbf{s}}_i)\le \epsilon_{\mathrm{RMS}}$ it suffices to choose
$\epsilon_{\mathrm{ST}}=\epsilon_{\mathrm{RMS}}$.

Using \cref{eq:N_ST} with $\delta\to\delta_t=\delta/N$ (so that $\log(M/\delta_t)=\log(NM/\delta)$), the number of shots per timestep is
\begin{align}
    N_{\mathrm{ST}} = \mathcal{O}\left(\frac{3^w}{\epsilon_{\mathrm{RMS}}^2}\log\frac{NM}{\delta}\right),
\end{align}
and the total number of ST shots across all $N$ timesteps is
\begin{align}
    \label{eq:N^ST_tot}
    N_{\mathrm{tot}}^{\mathrm{ST}}
    = N\cdot N_{\mathrm{ST}}
    = \mathcal{O}\left(\frac{3^w N}{\epsilon_{\mathrm{RMS}}^2}\log\frac{NM}{\delta}\right).
\end{align}
This number of shots suffices to guarantee $\mathrm{RMSE}(\mathbf{s}_i,\widehat{\mathbf{s}}_i)\le \epsilon_{\mathrm{RMS}}$ for all $i\in\{1,\dots,M\}$ with probability at least $1-\delta$.

Finally, we remark on our choice of baseline.
We use ``local classical shadows at all $N$ timesteps'' as a baseline because it is a standard, fully black-box
primitive for simultaneously estimating many Pauli observables from randomized measurements, and it matches the
``measure first, ask later'' setting where the observable family may be large.
Our time-subsampling idea is orthogonal to the per-timestep measurement primitive: in particular, it can be
combined with improvements such as Pauli grouping, derandomized or biased shadow ensembles, or other measurement
allocation strategies that reduce per-timestep variance.
At the level of our bounds, such improvements correspond to replacing the per-timestep shadow cost/variance proxy
by that of the chosen primitive, while the time-subsampling and compressed-sensing scaling in $m$ is unchanged.

\subsection{Compressed sensing from randomly sampled timesteps}

We now show how to reconstruct the same signal using compressed sensing.
Let us fix an observable index $i$ and suppress it for simplicity (i.e., write $\mathbf{s}$ for $\mathbf{s}_i$ and $\widehat{\mathbf{s}}$ for $\widehat{\mathbf{s}}_i$).
Let $\mathbf{s}\in\mathbb{R}^N$ denote the length-$N$ signal we wish to reconstruct.
Let $F\in\mathbb{C}^{N\times N}$ be a unitary transform (e.g., DFT or an orthonormal DCT) and define the coefficient vector
$\mathbf{x} \equiv F\mathbf{s}$.
Assume $\mathbf{x}$ is (approximately) $s$-sparse.\footnote{A vector $\mathbf{x}\in\mathbb{C}^N$ is \emph{$s$-sparse} if it has at most $s$ nonzero entries, i.e., $\|\mathbf{x}\|_0 \equiv \bigl|\{j\in\{1,\dots,N\}:\ x_j\neq 0\}\bigr| \le s$. Let $\mathbf{x}_s$ denote the vector obtained from $\mathbf{x}$ by retaining its $s$ largest-magnitude entries (and setting the rest to zero). We say that $\mathbf{x}$ is \emph{approximately $s$-sparse} if the tail $\|\mathbf{x}-\mathbf{x}_s\|_1$ is small. See \cref{app:cs} for more details.}

We first define the Restricted Isometry Property (RIP), which plays a key role in compressed sensing.

\begin{definition}[Restricted Isometry Property \cite{candesDecodingLinearProgramming2005}]
    \label{def:rip}
    For each integer $s=1,2,\dots$, define the isometry constant $\delta_s$ of a matrix $A$ as the smallest number such that
    \begin{align}
    \label{eq:RIP}
    (1-\delta_s)\|\mathbf{x}\|_2^2 \le \|A\mathbf{x}\|_2^2 \le (1+\delta_s)\|\mathbf{x}\|_2^2
    \end{align}
    holds for all $s$-sparse vectors $\mathbf{x}$. If this holds, then $A$ is said to obey the RIP of order $s$ with constant $\delta_s$.
\end{definition}

Let $\Omega\subset\{1,\dots,N\}$ be a uniformly random subset of size $m$. Throughout, $m$ denotes the number of sampled timesteps at which ST is performed in the CSST protocol.
Let $P_\Omega\in\{0,1\}^{m\times N}$ be the row-selection matrix that extracts entries of a length-$N$ vector on indices in $\Omega$.
For compatibility with the RIP, it is convenient to work with the normalized sampling operator\footnote{The factor $\sqrt{N/m}$ ensures the normalization used in standard RIP analyses for subsampled unitary matrices: for unitary $F$ one has $\mathbb{E}_\Omega\|(\sqrt{N/m} P_\Omega F^\dagger)\mathbf{x}\|_2^2=\|\mathbf{x}\|_2^2$. See \cref{app:cs} for details.}
\begin{align}
    \label{eq:R_Omega}
    R_\Omega \equiv \sqrt{\frac{N}{m}} P_\Omega .
\end{align}
In our setting, the sampled entries $(P_\Omega \mathbf{s}_i)_j$ are estimated by shadow tomography.
Let $\mathbf{e}_i^{\rm raw}\in\mathbb{R}^m$ denote the corresponding (unscaled) estimation error, so that
\beq
    \widehat{\mathbf{s}}_{i,\Omega} \equiv P_\Omega \mathbf{s}_i + \mathbf{e}_i^{\rm raw}.
\eeq
We apply the normalization
in classical post-processing and define the rescaled measurement vector
\beq
    \mathbf{y}_i \equiv \sqrt{\frac{N}{m}} \widehat{\mathbf{s}}_{i,\Omega}
    = R_\Omega \mathbf{s}_i + \mathbf{e}_i,\quad
    \mathbf{e}_i \equiv \sqrt{\frac{N}{m}} \mathbf{e}_i^{\rm raw}.
\eeq
With $\mathbf{x}_i \equiv F\mathbf{s}_i$ and $A \equiv R_\Omega F^\dagger$, this yields the (rescaled) measurement model
\begin{align}
    \label{eq:meas-model}
    \mathbf{y}_i = A\mathbf{x}_i + \mathbf{e}_i.
\end{align}
If shadow tomography guarantees the entry-wise bound $\|\mathbf{e}_i^{\rm raw}\|_\infty \le \epsilon_{\mathrm{ST}}$, then 

\begin{align}
    \label{eq:cs_e_ell2}
    \|\mathbf{e}_i\|_2 \le \sqrt{N} \epsilon_{\mathrm{ST}}.
\end{align}

Because the sensing operator is normalized by $\sqrt{N/m}$, each sampled entry is effectively upweighted.
Thus, at this worst-case level of bounding, the overall $\ell_2$ noise radius is $\eta=\Theta(\sqrt{N}\epsilon_{\mathrm{ST}})$
and does not decrease with $m$; the shot savings come from measuring only $m\ll N$ timesteps, not from relaxing the
per-timestep shadow accuracy requirement.

We will use the following standard stability guarantee for $\ell_1$ minimization, applied with a conservative noise radius $\eta$ satisfying $\|\mathbf{e}\|_2\le \eta$.

\begin{theorem}[Compressed sensing {\cite{candesStableSignalRecovery2005, candesIntroductionCompressiveSampling2008}}]
    \label{th:cs}
    Let $\mathbf{x}\in\mathbb{C}^N$ and measurements $\mathbf{y}=A\mathbf{x}+\mathbf{e}$ with $\|\mathbf{e}\|_2\le \eta$.
    Assume $A$ satisfies the RIP of order $2s$ with constant $\delta_{2s}<4/\sqrt{41}$.
    Let $\mathbf{x}^*$ solve the $\ell_1$-constrained least squares problem
    \begin{align}
        \mathbf{x}^*\in\arg\min_{\mathbf{z}}\|\mathbf{z}\|_1 \quad \text{s.t.}  \quad \|\mathbf{y}-A\mathbf{z}\|_2\le \eta .
    \end{align}
    Then
    \begin{align}
        \|\mathbf{x}-\mathbf{x}^*\|_2 \le c_1 \frac{\|\mathbf{x}-\mathbf{x}_s\|_1}{\sqrt{s}} + c_2 \eta,
    \end{align}
    where $\mathbf{x}_s$ retains the $s$ largest (in magnitude) entries of $\mathbf{x}$ and $c_1,c_2$ depend only on the RIP constant (hence are absolute constants once $\delta_{2s}$ is fixed below the threshold).
\end{theorem}

Since $F$ is unitary and the $\ell_2$ norm is unitarily invariant, $\|\mathbf{s}-\mathbf{s}^*\|_2=\|\mathbf{x}-\mathbf{x}^*\|_2$ where $\mathbf{s}^*\equiv F^\dagger \mathbf{x}^*$.
Using \cref{th:cs} with \cref{eq:cs_e_ell2} (so $\eta=\sqrt{N}\epsilon_{\mathrm{ST}}$) gives the error decomposition
\begin{align}
    \label{eq:cs_error}
    \|\mathbf{s}-\mathbf{s}^*\|_2
    \le
    c_1 \frac{\|F\mathbf{s}-(F\mathbf{s})_s\|_1}{\sqrt{s}} + c_2 \sqrt{N} \epsilon_{\mathrm{ST}}.
\end{align}
The first term is a model/compressibility error (zero when $F\mathbf{s}$ is exactly $s$-sparse),
and the second term quantifies how shadow-estimation noise propagates through the CS reconstruction.

\subsection{How large must $m$ be?}

To connect our sampling operator $R_\Omega$ with the standard compressed sensing measurement model in \cref{th:cs}, recall that we parametrize the time-domain signal $\mathbf{s}\in\mathbb{R}^N$ by its coefficients $\mathbf{x}\in\mathbb{C}^N$ in an orthonormal transform basis $F$ (e.g., DFT/DCT), i.e.,
$\mathbf{x} \equiv F\mathbf{s}$.
Our data consist of $m$ (noisy) samples of $\mathbf{s}$ at the randomly chosen timesteps $\Omega$, collected by the normalized row-selector
$R_\Omega \equiv \sqrt{\frac{N}{m}} P_\Omega \in \mathbb{R}^{m\times N}$.
Thus the rescaled measurement vector is
$\mathbf{y} = R_\Omega \mathbf{s} + \mathbf{e}
          = R_\Omega F^\dagger \mathbf{x} + \mathbf{e}$.
In other words, in the notation of \cref{th:cs} the measurement matrix is exactly the composition of ``synthesis'' (the transform $F^\dagger$) followed by ``subsampling'' ($m\ll N$ random samples): $A \equiv R_\Omega F^\dagger \in \mathbb{C}^{m\times N}$, so that $\mathbf{y}=A\mathbf{x}+\mathbf{e}$ [\cref{eq:meas-model}].
When applying \cref{th:cs} in our setting, we therefore take $\|\mathbf{e}\|_2 \le \eta = \sqrt{N} \epsilon_{\mathrm{ST}}$ [bounded as in \cref{eq:cs_e_ell2}].

To ensure \cref{th:cs} applies, we require $A=R_\Omega F^\dagger$ to satisfy the RIP of order $2s$.
The following standard result for subsampled bounded-unitary systems implies this holds with high probability for $m$ scaling nearly linearly in $s$.

\begin{theorem}[Fourier Compressed Sensing {\cite{havivRestrictedIsometryProperty2015}}]
    \label{th:fcs}
    Let $V\in\mathbb{C}^{N\times N}$ be unitary and assume $\|V\|_\infty \le C/\sqrt{N}$.
    Fix an integer $k\ge 1$ and $\delta_k\in(0,1)$.
    Let $A\in\mathbb{C}^{m\times N}$ be formed by sampling $m$ rows of $V$ uniformly at random, multiplied by $\sqrt{N/m}$.
    Then for
    \begin{align}
        m = \mathcal{O}\left(\delta_k^{-4} k \log^2\Big(\frac{k}{\delta_k}\Big) \log N\right),
    \end{align}
    the matrix $A$ satisfies RIP of order $k$ with constant $\delta_k$ with probability $1 - 2^{-\Omega (\delta_k^{-2} \log{N} \log{(k/\delta_k)})}$.
\end{theorem}

We apply \cref{th:fcs} with $V=F^\dagger$ (noting that $\|F^\dagger\|_\infty=\|F\|_\infty$), with $k=2s$, and choose $\delta_{2s}=4/\sqrt{41}-\kappa$ for any fixed $0<\kappa\ll 1$.
This yields the sampling rate
\begin{align}
    \label{eq:m_scaling}
    m = \mathcal{O}\left(s\log^2 s  \log N\right),
\end{align}
under which \cref{eq:cs_error} holds with high probability over the random choice of the sampled index set $\Omega$ (i.e., the sampled rows).

To obtain an overall failure probability at most $\delta$, we split the budget as
$\delta=\delta_{\mathrm{ST}}+\delta_{\mathrm{RIP}}$ and apply a union bound over the two independent sources of randomness: (i) the randomized measurements used by shadow tomography and (ii) the random choice of the sampled timestep set $\Omega$ which controls whether $A=\sqrt{N/m} P_\Omega F^\dagger$ satisfies RIP.  Throughout, we take $\delta_{\mathrm{ST}}=\delta_{\mathrm{RIP}}=\delta/2$ for simplicity.
\cref{th:fcs} provides a tail bound of the form
$\Pr(\text{$A$ fails RIP of order $k$}) \le 2^{-\Omega(\delta_k^{-2}\log N\log(k/\delta_k))}$.
Thus, for any desired $\delta_{\mathrm{RIP}}\in(0,1)$, it suffices to choose the implicit constant in the sampling rate \cref{eq:m_scaling} large enough so that this failure probability is at most $\delta_{\mathrm{RIP}}$.  In our asymptotic discussion we suppress this dependence and treat it as absorbed into the polylogarithmic factors.

\subsection{Exact sparsity: end-to-end guarantee, shot complexity, and shot reduction}
\label{sec:exact-sparse}

We formally summarize our results so far as follows:

\begin{theorem}[End-to-end guarantee assuming exact sparsity]\label{thm:end_to_end_rmse}
    Fix a total failure probability $\delta\in(0,1)$ and an RMSE target $\epsilon_{\mathrm{RMS}}>0$.
    Let $\delta_{\mathrm{ST}}=\delta/2$ and $\delta_{\mathrm{RIP}}=\delta/2$.
    Let $\Omega\subset\{1,\dots,N\}$ be a uniformly random subset of size $m$ (the sampled timesteps), and let
    $F\in\mathbb{C}^{N\times N}$ be an orthonormal transform (DFT or orthonormal DCT).
    Let $P_\Omega\in\{0,1\}^{m\times N}$ denote the row-selection matrix that extracts entries indexed by $\Omega$, and define the
    (normalized) measurement matrix
    \begin{equation}
        A  =  \sqrt{\frac{N}{m}} P_\Omega F^\dagger.
    \end{equation}
    For each observable $i\in\{1,\dots,M\}$, let $\mathbf{s}_i\in\mathbb{R}^N$ denote the length-$N$ signal with entries
    $(\mathbf{s}_i)_j = S_{ij}$, and define the coefficient vector $\mathbf{x}_i \equiv F\mathbf{s}_i$.
    Assume \emph{exact sparsity}: for every $i\in\{1,\dots,M\}$, $\mathbf{x}_i$ is $s$-sparse.
    
    At each sampled timestep $j\in\Omega$, perform $N_{\mathrm{ST}}$ Pauli-shadow snapshots and form estimates $\widehat{S}_{ij}$ of $S_{ij}$ for all $i\in\{1,\dots,M\}$.
    For each observable index $i\in\{1,\dots,M\}$, let $\widehat{\mathbf{s}}_{i,\Omega}\in\mathbb{R}^m$ denote the vector of shadow
    estimates at the sampled timesteps, i.e., if $\Omega=\{j_1,\dots,j_m\}$ (with $j_1<\cdots<j_m$ and $P_\Omega$ extracts entries in that same order) then
    $(\widehat{\mathbf{s}}_{i,\Omega})_k = \widehat{S}_{i,j_k}$ with $(k=1,\dots,m)$.
    Define the rescaled measurement and noise vectors
    \begin{equation}
        \mathbf{y}_i  \equiv  \sqrt{\frac{N}{m}} \widehat{\mathbf{s}}_{i,\Omega},\quad 
        \mathbf{e}_i  \equiv  \sqrt{\frac{N}{m}}\Bigl(\widehat{\mathbf{s}}_{i,\Omega}-P_\Omega \mathbf{s}_i\Bigr),
    \end{equation}
    so that $\mathbf{y}_i = A\mathbf{x}_i + \mathbf{e}_i$.
    
    If $N_{\mathrm{ST}}$ is chosen so that, conditioned on the sampled set $\Omega$, with probability at least $1-\delta_{\mathrm{ST}}$,
    \begin{equation}
        \max_{i\in\{1,\dots,M\}} \max_{j\in\Omega}  \bigl|\widehat{S}_{ij}-S_{ij}\bigr|  \le  \epsilon_{\mathrm{ST}},
    \end{equation}
    and $m$ is chosen so that, with probability at least $1-\delta_{\mathrm{RIP}}$, the matrix $A$ satisfies the RIP of order $2s$
    with constant $\delta_{2s}<4/\sqrt{41}$, then for every $i$ we have $\|\mathbf{e}_i\|_2 \le \eta$ with $\eta  \equiv  \sqrt{N} \epsilon_{\mathrm{ST}}$.
    Consequently, the Quadratically Constrained Basis Pursuit (QCBP) reconstruction
    \begin{equation}
        \label{eq:QCBP}
        \widehat{\mathbf{x}}_i \in \arg\min_{z\in\mathbb{C}^N}\|\mathbf{z}\|_1\quad \text{s.t.}\quad \|A\mathbf{z}-\mathbf{y}_i\|_2\le \eta,
        \quad \widehat{\mathbf{s}}_i \equiv F^\dagger \widehat{\mathbf{x}}_i,
    \end{equation}
    obeys, for all $i\in\{1,\dots,M\}$,
    \begin{equation}
        \mathrm{RMSE}(\mathbf{s}_i,\widehat{\mathbf{s}}_i) \le c_2 \epsilon_{\mathrm{ST}},
        \label{eq:rmse_end_to_end_exact}
    \end{equation}
    with probability at least $1-\delta$ over the random choice of $\Omega$ and all measurement outcomes.
    Here $c_2>0$ depends only on the RIP constant. In particular, choosing $\epsilon_{\mathrm{ST}}=\epsilon_{\mathrm{RMS}}/c_2$ achieves
    $\mathrm{RMSE}(\mathbf{s}_i,\widehat{\mathbf{s}}_i)\le \epsilon_{\mathrm{RMS}}$ for all $i\in\{1,\dots,M\}$.
\end{theorem}
Under exact sparsity, the model-mismatch term in \cref{eq:cs_error} vanishes, and \cref{thm:end_to_end_rmse} shows that achieving the target RMSE $\epsilon_{\mathrm{RMS}}$ reduces (up to constants) to ensuring per-sampled-timestep shadow accuracy $\epsilon_{\mathrm{ST}}=\Theta(\epsilon_{\mathrm{RMS}})$.

At each of the $m$ sampled timesteps, applying \cref{th:st} (absorbing the constant $c_2$ into the target accuracy) with per-timestep failure probability $\delta_t=\delta_{\mathrm{ST}}/m=\delta/(2m)$ gives
\begin{align}
    N_{\mathrm{ST}}
    = \mathcal{O}\left(\frac{3^w}{\epsilon_{\mathrm{RMS}}^2}\log\frac{mM}{\delta}\right),
\end{align}
so the total number of measurement shots for the CSST protocol (measuring at $m$ timesteps instead of all $N$) is
\begin{align}
    N_{\mathrm{tot}}^{\mathrm{CS}}
    = m\cdot N_{\mathrm{ST}}
    = \mathcal{O}\left(\frac{3^w m}{\epsilon_{\mathrm{RMS}}^2}\log\frac{mM}{\delta}\right).
\end{align}
Comparing with the baseline \cref{eq:N^ST_tot} yields the shot-count ratio
\begin{align}
    \label{eq:rf_theory}
    \frac{N_{\mathrm{tot}}^{\mathrm{ST}}}{N_{\mathrm{tot}}^{\mathrm{CS}}}
    =
    \Theta\left(
    \frac{N}{m}\cdot
    \frac{\log(NM/\delta)}{\log(mM/\delta)}
    \right),
\end{align}
up to absolute constant factors. See \cref{app:asymp_shots} for the asymptotic-notation conventions we use when comparing sufficient shot budgets.

If different observables have different effective sparsities, one may conservatively take $s=s_{\max}$, the largest effective sparsity among the $M$ signals.
Since $m=\mathcal{O}(s\log^2 s\log N) = \widetilde{\mathcal{O}}(s)$ [\cref{eq:m_scaling}], the reduction factor scales like $\widetilde{\Theta}(N/s)$, which justifies \cref{eq:shot-ratio}.
The advantage comes entirely from measuring only $m\ll N$ timesteps; in this worst-case analysis, the required per-timestep shadow accuracy is the same order as for the baseline at a fixed RMSE target.

\subsection{Approximate sparsity: compressibility for damped oscillations}
\label{sec:approx-sparse}

In many physical settings, observable time series are well approximated by a superposition of a modest number of oscillatory (and possibly damped) modes, which leads to sparse or rapidly decaying transform coefficients in a DFT/DCT domain.  In our numerical experiments (see \cref{sec:numerics}) we observe substantially faster coefficient decay than the worst-case analytic bound derived below.

Accordingly, we distinguish two viewpoints:
(i) a generic compressibility model (power-law decay of sorted coefficients), which matches typical empirical
behavior and yields standard compressed sensing error scaling; and
(ii) a signal-specific bound which is rigorous but intentionally
conservative and not tight in the parameter regimes we study numerically. 

In this section we pursue (ii) and conservatively bound the compressibility term $\|F\mathbf{s}-(F\mathbf{s})_s\|_1$ in \cref{eq:cs_error}, where $F$ is a unitary discrete Fourier- or cosine-type transform (DFT or DCT).

As a representative case, consider a single component sampled on $j=0,\dots,N-1$:
\begin{align}
    s_j = e^{-\gamma j\Delta t}\cos(\omega j\Delta t+\phi).
\end{align}
Let $F$ be the orthonormal DCT-II matrix.\footnote{The (type-II) discrete cosine transform (DCT-II) is the orthonormal linear map $F\in\mathbb{R}^{N\times N}$ with entries
$F_{kj}
=\sqrt{\frac{2-\delta_{k0}}{N}} 
\cos\left(\frac{\pi}{N}\Bigl(j+\tfrac12\Bigr)k\right)$, where $0\le j,k\le N-1$.} 
We show in \cref{app:approx-sparse} that then the DCT coefficients $\mathbf{x}=F\mathbf{s}$ ($\|\mathbf{x}\|_2=\|\mathbf{s}\|_2$) obey a Lorentzian-type envelope around the closest grid frequency $\nu_\ell=\pi\ell/T$ (with $T=N\Delta t$), leading to the tail bound
\begin{align}
    \label{eq:l1_tail_bound-0}
    \|\mathbf{x}-\mathbf{x}_s\|_1 \le \mathcal{O}\left(\sqrt{N} \log\frac{N}{s}\right),
\end{align}
where the hidden constant depends on $(\gamma,\omega,\Delta t)$ through the width of the corresponding spectral peak. Therefore, from \cref{eq:cs_error},
\begin{align}
    \label{eq:s-s*}
    \|\mathbf{s}-\mathbf{s}^*\|_2
    \le
    c_1\sqrt{\frac{N}{s}}\log\frac{N}{s} + c_2 \sqrt{N}\epsilon_{\mathrm{ST}}.
\end{align}

For a signal $s_j=\sum_{r=1}^{N_c} e^{-\gamma_r j\Delta t}|c_r|\cos(\omega_r j\Delta t+\phi_r)$ that is a sum of $N_c$ damped oscillations, 
linearity still gives a tail bound of the form \cref{eq:l1_tail_bound-0}.
In the worst case this can degrade the constant in \cref{eq:l1_tail_bound-0} roughly
linearly in $N_c$; in practice, the effective sparsity is governed by the number of
well-separated peaks in the DCT domain and their widths (set by damping and by the finite
time window).

Let us next determine the impact of approximate sparsity on the shot count ratio, i.e., the generalization of \cref{eq:rf_theory}.
Dividing \cref{eq:s-s*} by $\sqrt{N}$ yields the RMSE bound
\begin{align}
    \mathrm{RMSE}(\mathbf{s},\mathbf{s}^*)
    \le
    B_{\mathrm{RMS}}(s) + c_2 \epsilon_{\mathrm{ST}},
    \quad
    B_{\mathrm{RMS}}(s)\equiv c_1\frac{1}{\sqrt{s}}\log\frac{N}{s}.
\end{align}
To guarantee a target accuracy $\mathrm{RMSE}(\mathbf{s},\mathbf{s}^*)\le \epsilon_{\mathrm{RMS}}$, it suffices to choose $\epsilon_{\mathrm{ST}}$ so that
$c_2 \epsilon_{\mathrm{ST}} \le \epsilon_{\mathrm{RMS}} - B_{\mathrm{RMS}}(s)$ (in particular, we require $B_{\mathrm{RMS}}(s) < \epsilon_{\mathrm{RMS}}$), i.e.,
\begin{align}
    \epsilon_{\mathrm{ST}}
    =
    \Theta \left(\epsilon_{\mathrm{RMS}} - B_{\mathrm{RMS}}(s)\right).
\end{align}
At each of the $m$ sampled timesteps, \cref{th:st} (with $\delta\to \delta_t=\delta/m$) gives
\begin{align}
    N_{\mathrm{ST}}
    =
    \mathcal{O}\left(
    \frac{3^w}{(\epsilon_{\mathrm{RMS}}-B_{\mathrm{RMS}}(s))^2}\log\frac{mM}{\delta}
    \right),
\end{align}
hence the total shot-count for the CS scheme is
\begin{align}
    N_{\mathrm{tot}}^{\mathrm{CS}}
    =
    m N_{\mathrm{ST}}
    =
    \mathcal{O}\left(
    \frac{3^w m}{(\epsilon_{\mathrm{RMS}}-B_{\mathrm{RMS}}(s))^2}\log\frac{mM}{\delta}
    \right).
\end{align}
Meanwhile the baseline (measuring all $N$ timesteps, no CS) remains \cref{eq:N^ST_tot}.
Therefore, the shot-count ratio in the approximately sparse setting is
\begin{align}
    \label{eq:rf_theory-approx}
    \frac{N_{\mathrm{tot}}^{\mathrm{ST}}}{N_{\mathrm{tot}}^{\mathrm{CS}}}
    =\Theta \left(
    \frac{N}{m}\cdot
    \left(\frac{\epsilon_{\mathrm{RMS}}-B_{\mathrm{RMS}}(s)}{\epsilon_{\mathrm{RMS}}}\right)^2\cdot
    \frac{\log(NM/\delta)}{\log(mM/\delta)}
    \right),
\end{align}
up to absolute constant factors.

Note that if we choose $s$ large enough that $B_{\mathrm{RMS}}(s)\le \epsilon_{\mathrm{RMS}}/2$, then
$(\epsilon_{\mathrm{RMS}}-B_{\mathrm{RMS}}(s))/\epsilon_{\mathrm{RMS}}=\Theta(1)$ and \cref{eq:rf_theory-approx} reduces (up to constants) to the exact-sparsity scaling \cref{eq:rf_theory}. Note further that, as in \cref{eq:m_scaling}, to ensure RIP for the subsampled DFT/DCT case, we still take $m=\mathcal{O}(s\log^2 s\log N)$ (with $s$ now interpreted as the effective sparsity level used in the reconstruction); the additional requirement in the approximate case is that $s$ must be large enough so that $B_{\mathrm{RMS}}(s)$ is below the target tolerance.

Finally, let us briefly consider case (i) mentioned at the beginning of this section, i.e., a generic power-law compressibility model.
Let $\mathbf{x}=F\mathbf{s}$ and let $|\mathbf{x}|_{(1)}\ge |\mathbf{x}|_{(2)}\ge\cdots$ denote the coefficients sorted by magnitude.
Assume that for some $r>1$ and $C_r>0$,
\begin{equation}
    |\mathbf{x}|_{(k)} \le C_r k^{-r}\quad (k=1,2,\dots,N).
    \label{eq:powerlaw_compressibility}
    \end{equation}
    Then the best $s$-term approximation $\mathbf{x}_s$ obeys the standard tail bound
    \begin{equation}
    \|\textbf{x}-\mathbf{x}_s\|_1 \le \frac{C_r}{r-1} s^{1-r}.
    \label{eq:l1_tail_powerlaw}
\end{equation}
Combined with the QCBP stability guarantee, this yields a reconstruction error that decays polynomially in $s$.

\section{Numerical Results}
\label{sec:numerics}

In this section, we empirically study CSST on simulated open-system dynamics: we compute ``ground-truth'' expectation-value time series from the master equation and then simulate finite-shot Pauli-shadow data (random measurement bases and outcomes) to produce the ST and CSST estimates.  While the bounds in \cref{sec:csst_bounds} provide worst-case asymptotic guarantees in terms of an (often unknown) sparsity level $s$, the experimentally tunable parameters are (i) the number of sampled timesteps $m$ and (ii) the number of classical-shadow snapshots per sampled timestep $N_{\mathrm{ST}}$.  For each observable $O_i$ we report the reconstruction error using
\begin{align}
    \label{eq:RMSE}
    \mathrm{RMSE}(\mathbf{s}_i,\widehat{\mathbf{s}}_i)\equiv
    \sqrt{\frac{1}{N}\sum_{j=1}^N\bigl(S_{ij}-\widehat{S}_{ij}\bigr)^2},
\end{align}
and we typically average RMSE over all Pauli strings within a fixed weight sector.

Our theoretical guarantees are stated for the constrained $\ell_1$ QCBP reconstruction in \cref{eq:QCBP}.

In practice, we solve a Lagrangian (penalized) counterpart, namely the LASSO problem (implemented using scikit-learn's \verb|linear_model.Lasso| \cite{pedregosaScikitlearnMachineLearning}),
\beq
    \label{eq:lasso}
    \widehat{\mathbf{z}}(\alpha)\in \arg\min_{\mathbf{z}}
    \frac{1}{2N}\|A\mathbf{z}-\mathbf{y}\|_2^2+\alpha\|\mathbf{z}\|_1.
\eeq
Using the definitions $A=\sqrt{N/m}\,P_\Omega F^\dagger$ and $\mathbf{y}=\sqrt{N/m}\,\widehat{\mathbf{s}}_{i,\Omega}$, this can be written equivalently as
\beq
    \arg\min_{\mathbf{z}}
    \frac{1}{2m}\bigl\|P_\Omega F^\dagger \mathbf{z}-\widehat{\mathbf{s}}_{i,\Omega}\bigr\|_2^2+\alpha\|\mathbf{z}\|_1,
\eeq
which matches scikit-learn’s convention of placing a factor $1/(2m)$ in front of the squared loss.
(Alternative prefactors are purely conventional and only rescale the numerical value of $\alpha$ unless accompanied by a compensating rescaling of $A$ and $\mathbf{y}$.)
Rather than selecting $\alpha$ via a target residual constraint $\|A\mathbf{z}-\mathbf{y}\|_2\le \eta$, in our numerical experiments (where the ground truth is available) we perform an $\alpha$-sweep to quantify the best achievable performance of the penalized formulation for our measurement matrix $A$:
we vary $\alpha$ logarithmically from $10^{-7}$ to $10^{-2}$ and report the choice of $\alpha$ that minimizes the empirical RMSE.

\begin{figure}
    \centering
    \includegraphics[width=0.98\linewidth]{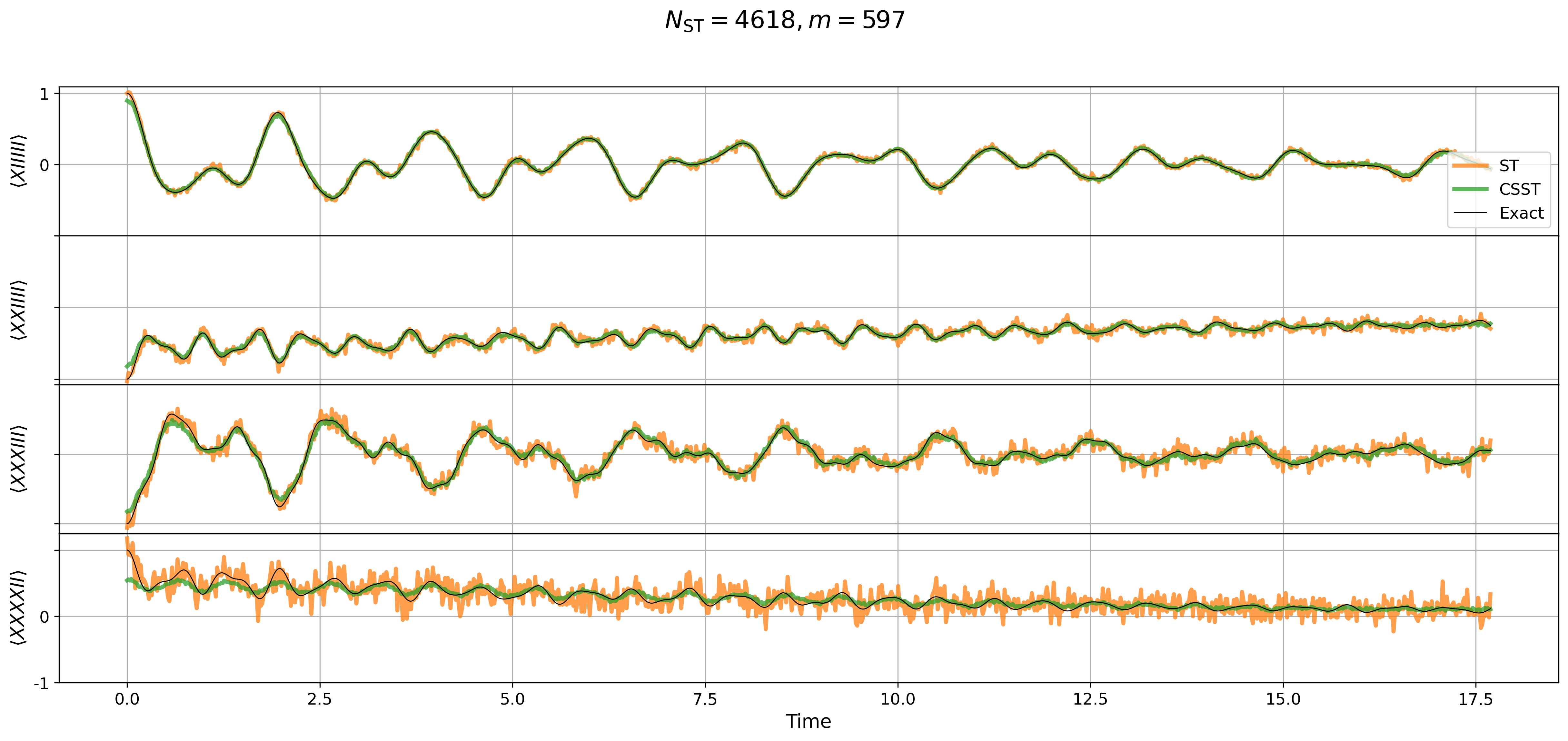}
    \caption{$\langle O(t)\rangle$ over time for four selected observables of the $2\times3$ noisy  Heisenberg model defined in \cref{eq:heis_ham,eq:lme}.  We plot the ground truth from exact simulation, the baseline estimation using only shadow tomography, and the CSST reconstructions at $\alpha^*$ from each observable's $\alpha$ sweep. Remarkably, the subsampled CSST reconstruction---which uses only 60\% of the available data---is significantly closer to the ground truth data than the baseline ST estimation.}
    \label{fig:cs_example}
\end{figure}

As a case study, we use QuTiP \cite{lambertQuTiP5Quantum2026} to simulate the Heisenberg model on $n=6$ qubits arranged on a $2\times 3$ rectangular lattice with open boundary conditions.  The Hamiltonian is
\begin{align}
    \label{eq:heis_ham}
    H = J\sum_{\langle a,b\rangle}\left(X_aX_b+Y_aY_b+Z_aZ_b\right),
\end{align}
where $\langle a,b\rangle$ ranges over nearest-neighbor edges and we set $J=1$ to fix units.  The initial state is $\ket{+-+-+-}$, where $\ket{\pm}$ are the $\pm 1$ eigenstates of $X$.  Open-system dynamics are generated by the Lindblad master equation
\begin{align}
    \label{eq:lme}
    \dot{\rho}=-i[H,\rho]+\sum_{q=1}^n\Big(\gamma_\phi \mathcal{D}[Z_q](\rho)+\gamma_1 \mathcal{D}[\sigma^-_q](\rho)\Big),
    \quad
    \mathcal{D}[L](\rho)=L\rho L^\dagger-\tfrac12\{L^\dagger L,\rho\},
\end{align}
with uniform single-qubit dephasing and amplitude damping rates $\gamma_\phi=\gamma_1=\gamma=10^{-1}$.  We sample $N=1000$ timesteps with $\Delta t\approx 0.024$ (total window $T=N\Delta t$).  At each sampled timestep we simulate Pauli-shadow snapshots (random single-qubit measurement bases and outcomes) and use them to estimate every Pauli string up to weight $4$ via the standard Pauli-shadow estimators.

To illustrate our results, in \cref{fig:cs_example} we select four observables of increasing Pauli weight and plot their true expectation values, baseline estimations, and CSST reconstructions over time.  The baseline scheme uses $N_{\text{ST}}=7437$ shadows per timestep, and the CSST scheme randomly samples only $m=597$ of these timesteps.  The CSST reconstructions are noticeably closer to their exact forms than the baseline curves despite having ignored around $40$\% of the available data.  We provide additional examples in different $(N_\text{ST},m)$ parameter regimes in \cref{app:examples}.

\begin{figure*}
    \centering
    \makebox[\linewidth][c]{%
      \includegraphics[width=0.4\linewidth]{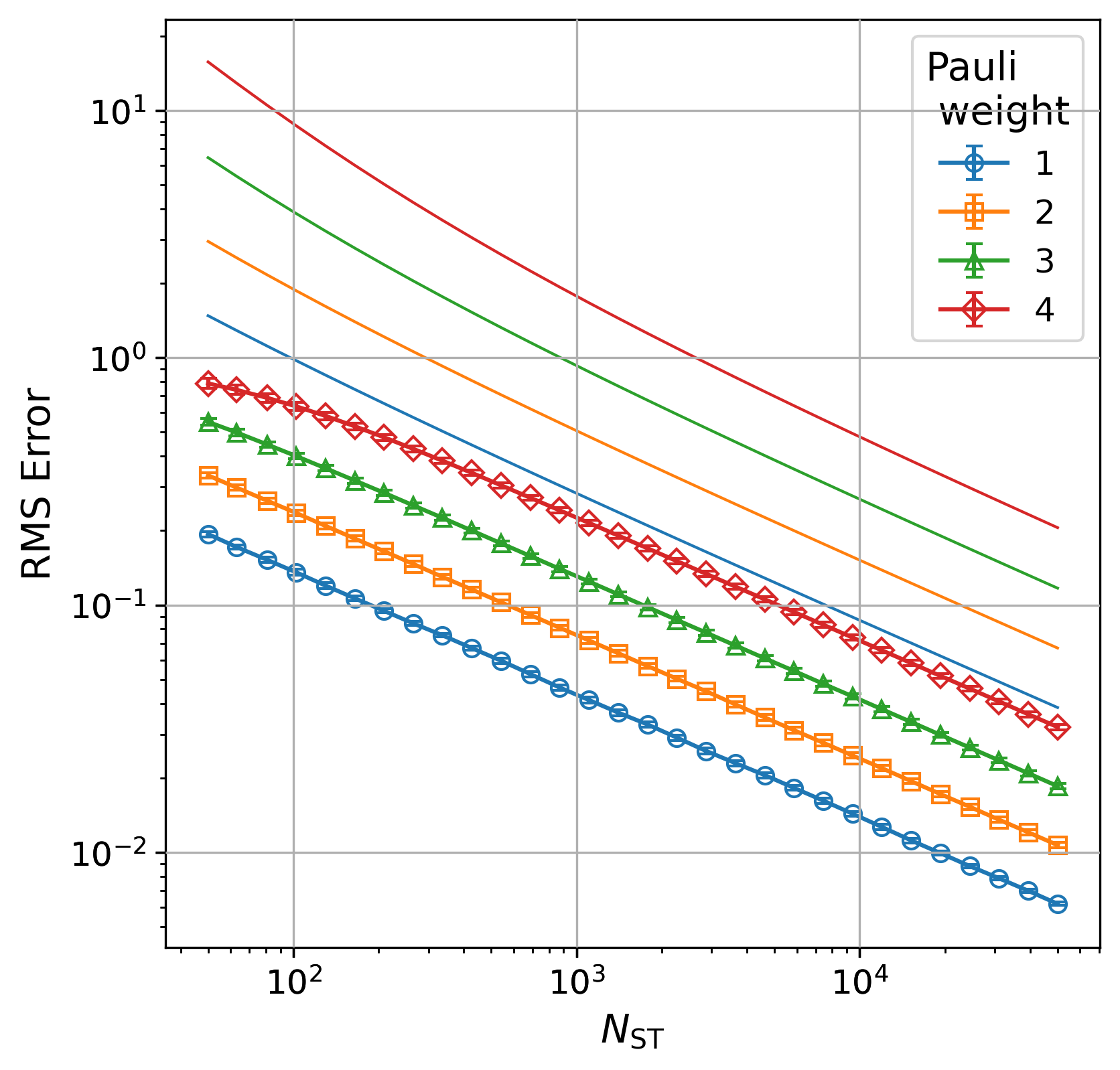}\hspace{0.05\linewidth}%
      \includegraphics[width=0.4\linewidth]{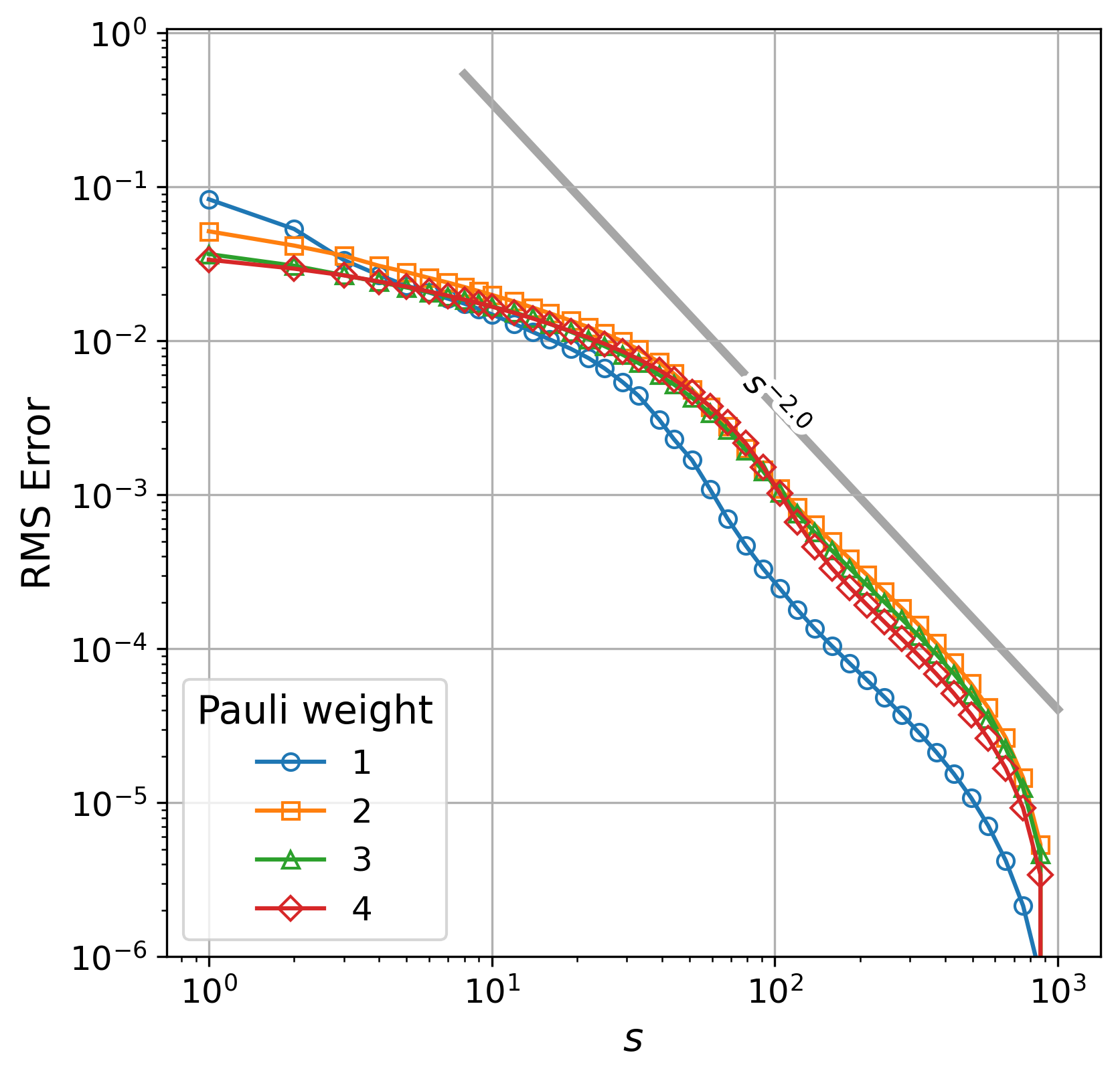}%
    }
    \caption{Baseline shadow-tomography noise levels and transform-domain compressibility for our noisy $2\times 3$ Heisenberg model.  Left: Baseline ST RMSE [\cref{eq:RMSE}] as a function of snapshots per timestep $N_{\mathrm{ST}}$, grouped by Pauli weight $w\in\{1,2,3,4\}$. Markers show empirical RMSEs averaged over all weight-$w$ Pauli strings; solid curves show the corresponding finite-sample upper bounds from the Pauli-shadow analysis using Bernstein's inequality [\cref{eq:bern}].  Right: Weight-sector-averaged truncation RMSE of the exact (noise-free) Pauli traces under the orthonormal DCT-II: for each trace we keep the $s$ largest-magnitude DCT coefficients, invert the transform, and compute the resulting RMSE. The dashed guideline indicates an approximate $\propto s^{-2}$ decay over the power-law regime ($s\gtrsim 10^2$), highlighting strong DCT-domain compressibility.}
    \label{fig:setup}
\end{figure*}

To understand these results and those that follow, in \cref{fig:setup} we simply compute the RMSE between each exact Pauli signal and its estimation using $N_{\text{ST}}$ shadows per timestep (left panel).  We hope to reduce this baseline RMSE via compressed sensing, even while using fewer measurements. 

We also inspect the compressibility of the exact signals in \cref{fig:setup} (right).  For each exact time series, we compute its orthonormal DCT-II coefficients,\footnote{The DCT is preferred to the DFT since our data is only defined for $t > 0$, so it makes no difference if we find solutions which correspond to even extensions of the data to $t < 0$; this simplification permits the use of real-valued computations throughout the reconstruction.} retain the $s$ largest-magnitude coefficients, and invert the transform; \cref{fig:setup} (right) reports the resulting truncation RMSE averaged within each Pauli-weight sector.  Over the range $10^{2}\lesssim s\lesssim 10^3$, the truncation RMSE decays approximately as a power law (guide line shown), indicating strong DCT-domain compressibility and motivating time-subsampled reconstruction. 

To interpret this power-law regime in \cref{fig:setup} (right), note that we are plotting the error of the best $s$-term approximation obtained by retaining the $s$ largest (in magnitude) transform coefficients. Let $x\in\mathbb{R}^N$ denote the DCT coefficient vector and let $\mathbf{x}_s$ be $\mathbf{x}$ restricted to its $s$ largest entries. A standard compressibility model assumes that the sorted magnitudes obey $|\mathbf{x}|_{(k)} \lesssim k^{-r}$ for some $r>1$, which implies the $\ell_1$ tail bound $\|\mathbf{x}-\mathbf{x}_s\|_1 = \mathcal{O}(s^{-r+1})$.  Since our plotted metric is an $\ell_2$-type error (RMSE), we instead control the tail:
\begin{equation}
    \|\mathbf{x}-\mathbf{x}_s\|_2^2  =  \sum_{k>s} |\mathbf{x}|_{(k)}^2  \lesssim  \sum_{k>s} k^{-2r}
     =  \mathcal{O}(s^{-2r+1}),
    \quad\text{hence}\quad
    \|\mathbf{x}-\mathbf{x}_s\|_2  =  \mathcal{O}(s^{-r+1/2}).
\end{equation}
Because the transform is orthonormal, truncation in coefficient space produces the same $\ell_2$ error in the signal domain, and therefore $\mathrm{RMSE}=\frac{1}{\sqrt{N}}\|\mathbf{x}-\mathbf{x}_s\|_2 = \mathcal{O}(s^{-r+1/2})$.  The empirical trend $\mathrm{RMSE}\propto s^{-2}$ is thus consistent with an approximate decay $|\mathbf{x}|_{(k)}\sim k^{-5/2}$ in the sorted coefficient magnitudes.

To study the scaling behavior of CSST, we vary the number of shadows per timestep ($N_{\mathrm{ST}}$) logarithmically from $10$ to $50000$.  Then, for each of those $N_{\mathrm{ST}}$ values, compressed sensing is run using $m$ randomly selected timesteps from $50$ to $1000$ (different curves within each panel), i.e., up to the maximum number of timesteps one would have used without compressed sensing.  
Additionally, at each pair $(N_{\mathrm{ST}},m)$ we sweep $\alpha$ on a logarithmic grid and solve \cref{eq:lasso}.  We again take $F$ to be the orthonormal DCT-II and use the normalized subsampled transform $A=\sqrt{\frac{N}{m}} P_\Omega F^T$.  The data vector $\mathbf{y}$ is the corresponding rescaled vector of the $m$ sampled time-domain estimates.

In what follows, we use two filters to preprocess the CSST datasets:
\begin{enumerate}
    \item We do not plot any data associated to any observable $O_i$ whose true signal's variance is less than $10^{-3}$.
    \item We do not plot any data associated to any observable-snapshot pair $(O_i, N_{\mathrm{ST}})$ whose signal-to-noise ratio (SNR) is below $1$.
\end{enumerate}
We compute the SNR \cite{oppenheimDiscretetimeSignalProcessing2014} via 
\begin{align}
    \label{eq:snr}
    \text{SNR}(O_i,N_{\mathrm{ST}}) = 10 \log_{10} \left( \frac{\sum_j S_{ij}^2}{\sum_j ( \hat{S}_{ij} - S_{ij})^2} \right) .
\end{align}
The motivation for these two conditions is as follows.  Working backwards from the intended use case of spectral estimation, signals which appear close to, or are exactly, constant do not contribute to interesting dynamical phenomena over the duration of observation, and thus are presumably of little interest to the observer.  Additionally, a very low $N_{\text{ST}}$ per timestep can result in signals which are too noisy to be recovered, in which case we may skip running CSST on them.  One can employ statistical tests such as the Ljung-box test to rigorously check whether a signal is statistically equivalent to white noise \cite{chanAlgorithmicShadowSpectroscopy2025, ljungMeasureLackFit1978} before processing it further, which does not require knowing the true signal unlike the SNR formula given in \cref{eq:snr}.  In our numerics, we use these checks primarily to reduce the computational overhead associated with scanning over many different $m$, $N_{\text{ST}}$, and $\alpha$ combinations, thereby filtering out in advance any such signals whose reconstructions would have likely failed.

\begin{figure}
    \centering
    \includegraphics[width=\linewidth]{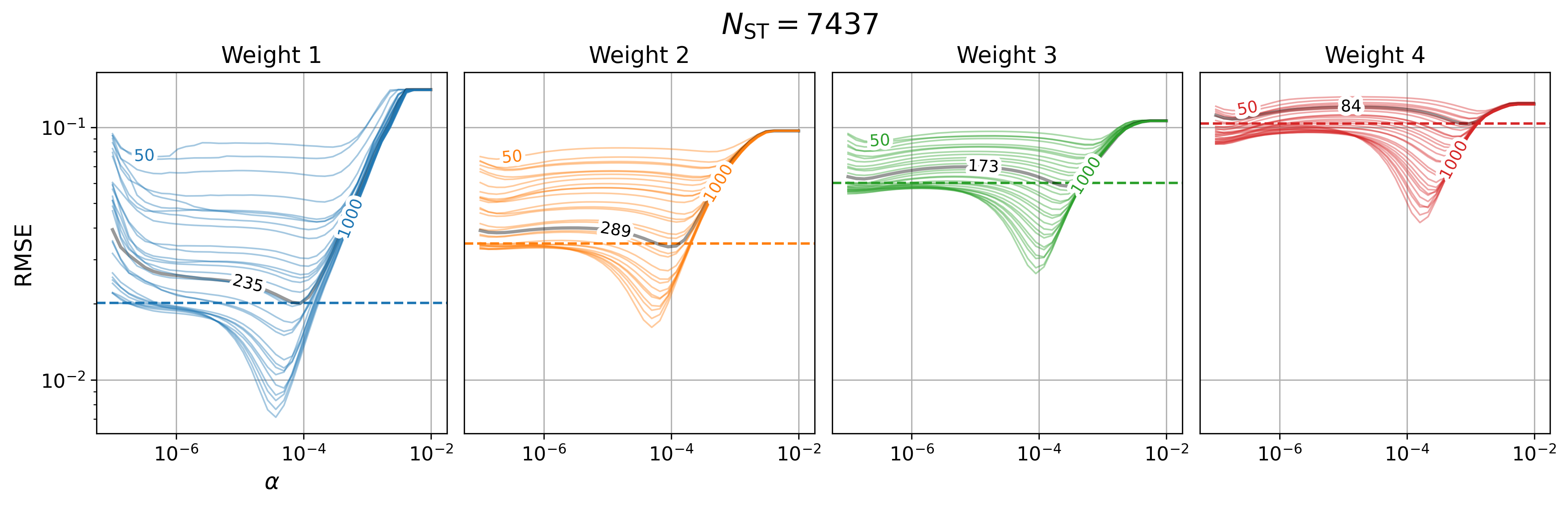}
    \caption{
    CSST reconstruction error versus LASSO regularization (same noisy $2\times 3$ Heisenberg model as in \cref{fig:setup}).
    Each panel corresponds to a Pauli-weight sector $w\in\{1,2,3,4\}$ and reports RMSE averaged over all weight-$w$ Pauli strings. Curves show the CSST reconstruction RMSE [\cref{eq:RMSE}] as a function of the regularization parameter $\alpha$ in \cref{eq:lasso}, for several choices of the number of randomly sampled timesteps $m$ (curve labels), at fixed snapshots $N_{\mathrm{ST}}=7437$ per sampled timestep.
    The dashed reference line indicates the average baseline ST RMSE using all $N$ timesteps without CS.
    Black annotations indicate $m^*$, the smallest sampled-timestep value for which the best-achieved RMSE (minimized over $\alpha$) beats the baseline ST line.}
    \label{fig:cserrs_sind}
\end{figure}

In \cref{fig:cserrs_sind}, for a fixed value of $N_{\mathrm{ST}}=7437$, we plot the RMSE of each Pauli signal reconstructed from some number $m$ of timesteps (see curve labeling), as a function of $\alpha$, averaged over all Pauli strings of a given weight (as labeled in the subplot titles).  The dashed line indicates the average RMSE achieved using all available timesteps without compressed sensing at all.  Any $m < 1000$ curve which dips below this threshold, at some $\alpha$, indicates that some timesteps can be neglected entirely without incurring an increase in error relative to the naive approach.  Denote the \textit{lowest} value of $m$ which satisfies this criterion by $m^{*}$ (these values are colored in black in each subplot).  Any larger value $m^{*} < m < 1000$ indicates a scenario where one can simultaneously measure \textit{less} often and still achieve \textit{lower} error.

\begin{figure*}
    \includegraphics[width=0.48\linewidth]{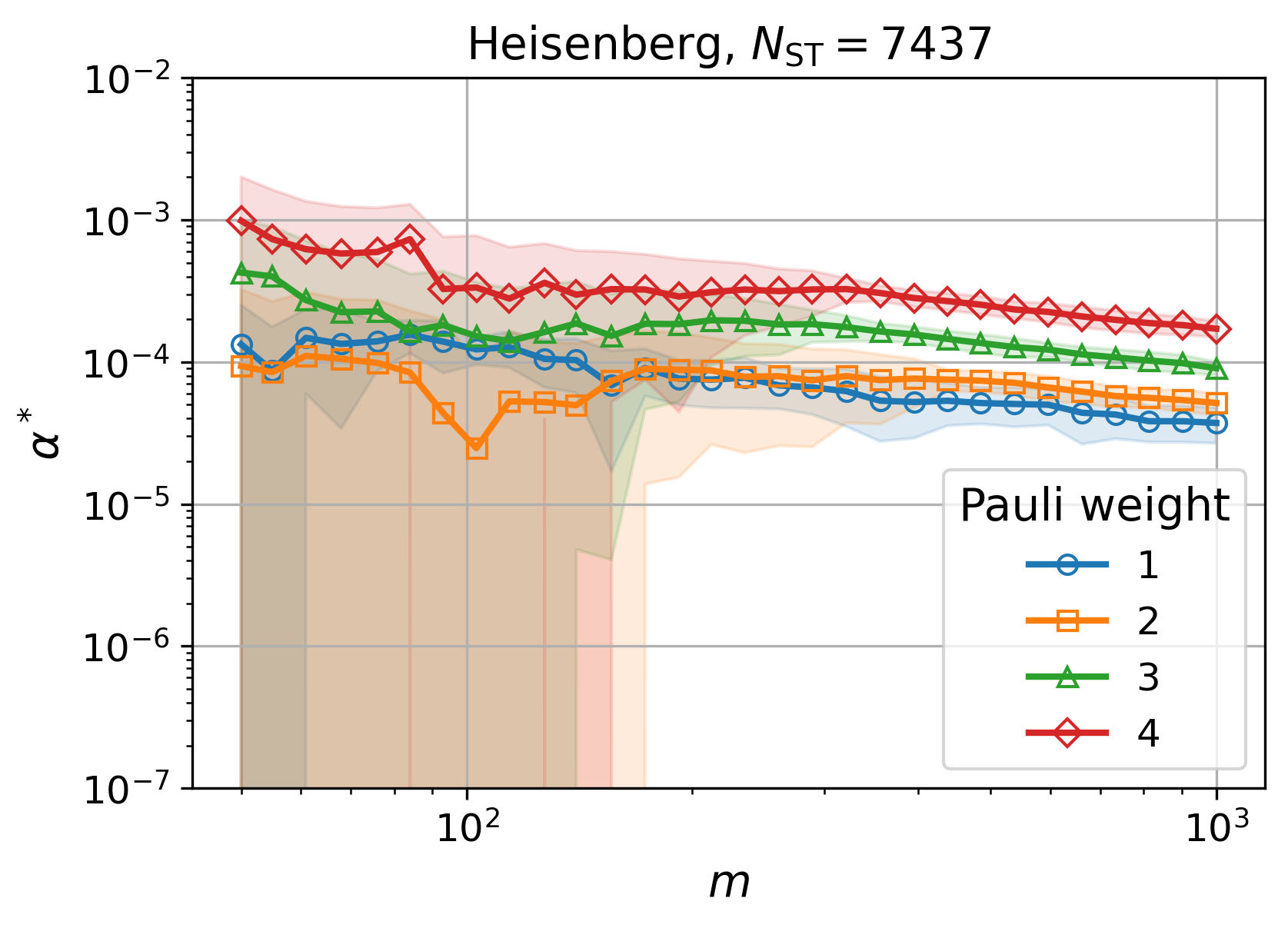}
    \hfill
    \includegraphics[width=0.48\linewidth]{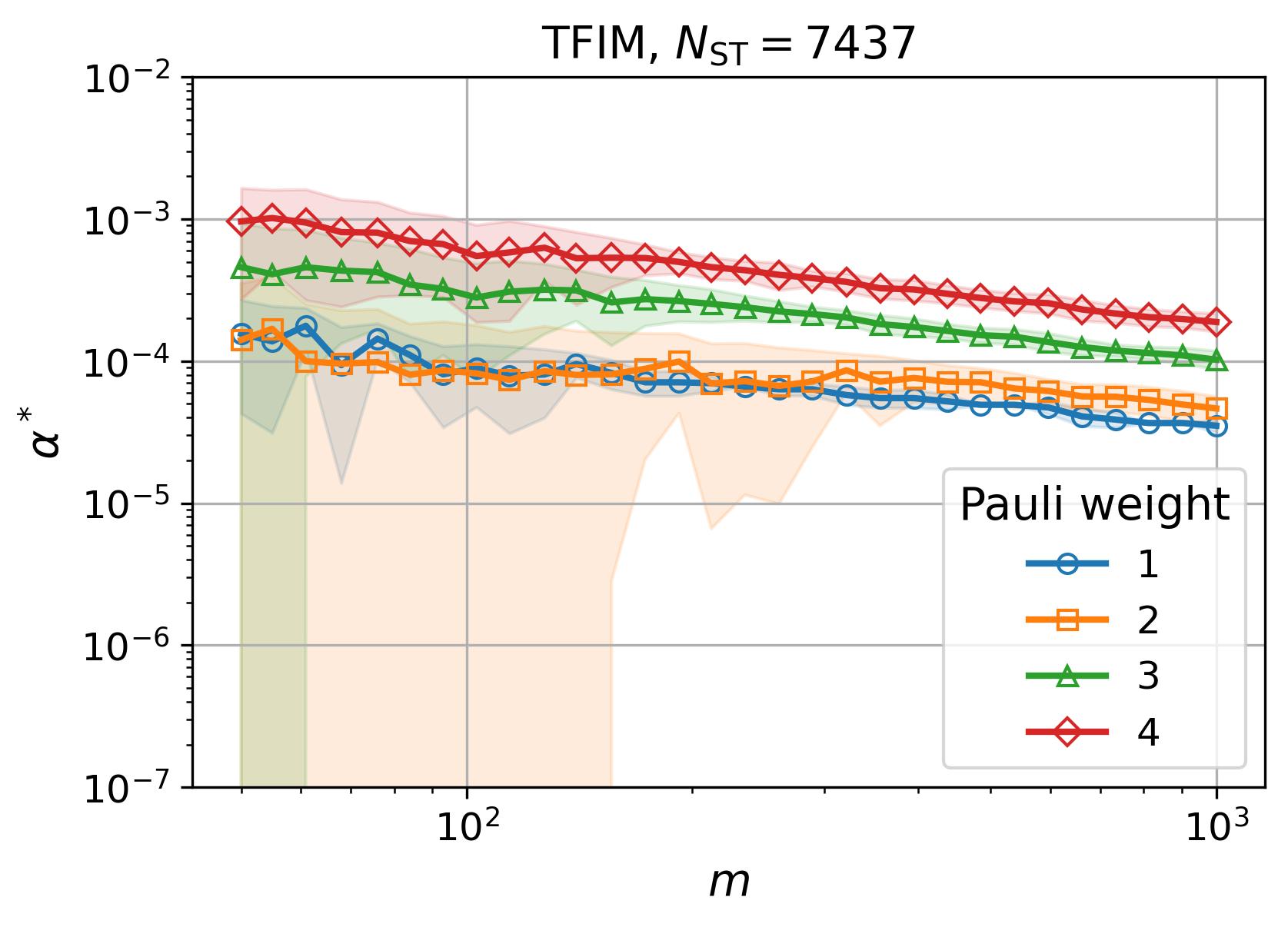}
    \caption{Optimal regularization parameters for CSST.  For each Pauli observable $O_i$ and each sampled-timestep budget $m$, we define the optimal value $\alpha_i^*$ as the value of $\alpha$ that minimizes the reconstruction RMSE over the $\alpha$-sweep shown in \cref{fig:cserrs_sind}. Lines show the mean of $\alpha_i^*$ within each Pauli-weight sector $w$ as a function of $m$, and shading shows one standard deviation across observables in that sector (same model and snapshot budgets as in \cref{fig:cserrs_sind}).  The stabilization (variance reduction) of $\alpha^*$ at larger $m$ suggests that (for fixed $w$) a single near-optimal regularization choice can work well across many observables.  Left: $\alpha_i^*$ for the $2\times3$ Heisenberg model.  Right: $\alpha_i^*$ for the $2\times3$ TFIM model with initial state $\ket{\text{GHZ}}$ (see \cref{app:tfim} for the model Hamiltonian and additional data).  Note the similarity with respect to the Heisenberg model, highlighting that the $\alpha_i^*$ are primarily a function of the noise only and not the intrinsic details of the chosen model.  Thus, an $\alpha^*$ which works well on average for a given Pauli weight from one model's dataset should be applicable to other models as well.  The sharp jump which occurs at low $m$ is due to the corresponding curves' minima in \cref{fig:cserrs_sind} moving from around $10^{-7}$ to $10^{-4}$.}
    \label{fig:optalphas}
\end{figure*}

While these results are promising, ideally there should be a way to choose $\alpha$ without first having to run exhaustive parameter sweeps.  It would not be very practical if the optimal $\alpha$ of each Pauli signal varied over a large range, in a potentially unpredictable way.\footnote{Note that here we first find the optimal $\alpha$ per Pauli signal, before averaging over Pauli weight sectors.  The curve minima in the \cref{fig:cserrs_sind} instead indicate the average optimal $\alpha$ per Pauli weight sector.}  For each Pauli signal (at each value of ($N_{\mathrm{ST}},m$)) we find the optimal $\alpha^*$ which minimizes the RMSE.  In \cref{fig:optalphas}, the signals are grouped by Pauli weight, and the means are plotted as a function of $m$ with error bars given by the standard deviation.  As $m$ increases, the average $\alpha^*$ stabilizes with low variance, indicating that good values for $\alpha^*$ may be chosen which work well across \textit{all} Pauli signals of a given weight.  These values tend to increase with decreasing $m$, since higher values of $\alpha$ are needed to impose more sparsity, and if a lower value of $m$ suffices to achieve a good reconstruction, then the true signal must also be sparse. 

\begin{figure}
    \centering
    \includegraphics[width=\linewidth]{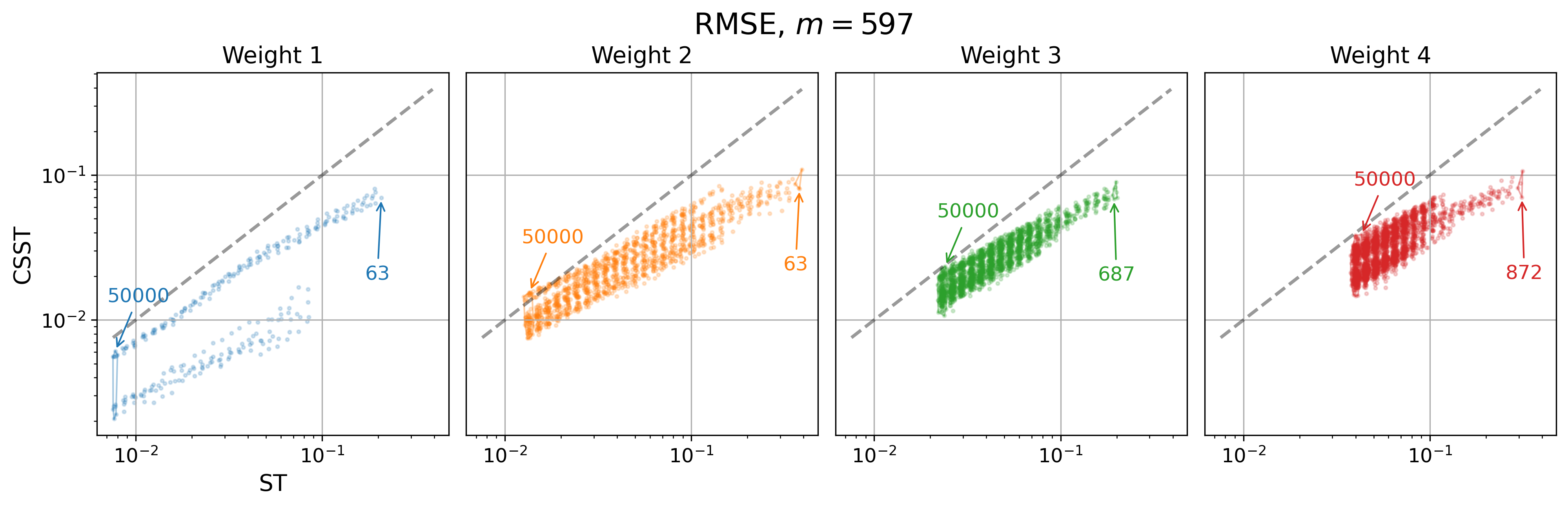}
    \caption{Per-observable comparison of baseline ST and CSST on the same noisy Heisenberg model as in the previous figures.  Each panel corresponds to a Pauli-weight sector $w\in\{1,2,3,4\}$, and each marker is a single weight-$w$ Pauli string $O_i$ whose length-$N$ expectation-value trace is estimated from finite-shot Pauli shadows and then (optionally) reconstructed via compressed sensing.
    Horizontal axis: the baseline ST reconstruction error, quantified by $\mathrm{RMSE}(\mathbf{s}_i,\widehat{\mathbf{s}}_i^{\mathrm{ST}})$ [\cref{eq:RMSE}] 
    obtained by measuring all $N$ timesteps with $N_{\mathrm{ST}}$ snapshots per timestep.
    Vertical axis: the best CSST reconstruction error quantified by $\min_{\alpha}\mathrm{RMSE}(\mathbf{s}_i,\widehat{\mathbf{s}}_i^{\mathrm{CS}}(\alpha))$ [\cref{eq:RMSE} again]
    obtained by measuring only a fixed subset of $m$ timesteps (here $m=0.6 N$, i.e., 60\% of the available times chosen uniformly at random without replacement) and reconstructing the full length-$N$ trace using \cref{eq:lasso}, with the regularization parameter $\alpha$ selected per observable to minimize RMSE.
    Different columns correspond to different snapshots-per-timestep budgets $N_{\mathrm{ST}}$ (values annotated); the convex hulls and arrows highlight, for visual aid, the spread between the smallest and largest $N_{\mathrm{ST}}$ used in each weight sector.
    The dashed line is the break-even line $y=x$: points below it indicate observables whose reconstruction error is {reduced} by CSST despite using only $m<N$ measured timesteps.}
    \label{fig:cserrs_tind}
\end{figure}

In \cref{fig:cserrs_tind}, we slice the data differently than in \cref{fig:cserrs_sind}.  We do not average within a Pauli weight sector; instead, each observable is plotted as its own data point whose $x$-value is the error achieved without compressed sensing, and whose $y$-value is the \textit{lowest} error achieved with compressed sensing on $m$ timesteps, at some optimal $\alpha$ (which varies between data points).  All data corresponding to the same $N_{\text{ST}}$ tends to fall in its own thin column since the ST errors have low variance across different Paulis (the largest and smallest values of $N_{\text{ST}}$ for each Pauli weight are annotated with arrows and wrapped in a thin convex hull for visual aid).  Compressed sensing offers a reduction in error for some Pauli signal if its data point lies below the black dashed line ($y=x$).  Generally, increasing Pauli weight or decreasing $N_{\text{ST}}$ both increase the error, which provides for leeway for compressed sensing to offer some benefit.  We see that in \cref{fig:cserrs_tind}, for weights $1,3,4$, nearly every observable benefits from compressed sensing even when restricting to only 60\% of the available data.  Some weight-$2$ observables do not benefit when the baseline error is small enough, but on average there is still an improvement.  The cluster of lowest $N_{\mathrm{ST}}$ varies between the Pauli weight subplots due to our pre-processing; specifically, the SNR condition allows us to skip the reconstruction of many weight $3$ and $4$ observables at low $N_{\text{ST}}$.

\begin{figure}
    \centering
    \includegraphics[width=\linewidth]{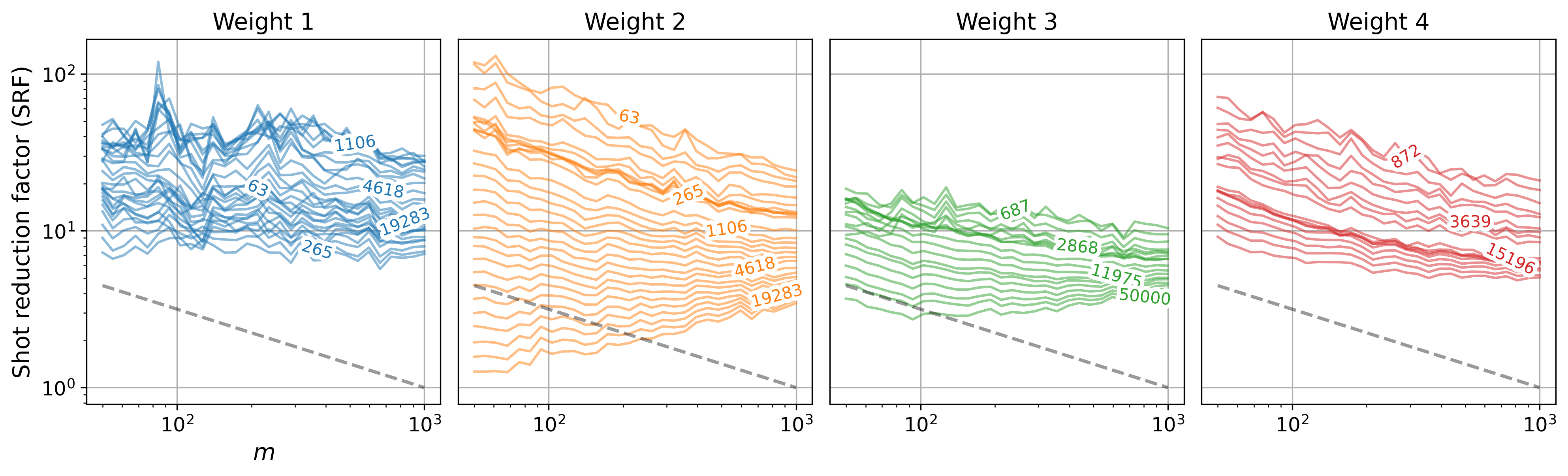}
    \caption{Weight-sector-averaged shot reduction factor (SRF) [\cref{eq:SRF}] inferred from the observed CSST error reductions on the same noisy Heisenberg model as in the previous figures.  Curves report $\mathrm{SRF}$ averaged over all Pauli strings in each weight sector as a function of the number of sampled timesteps $m$; each curve corresponds to a different per-timestep snapshot budget $N_{\mathrm{ST}}$ (annotated on the plot).  Values of $\mathrm{SRF}>1$ indicate a net reduction in the total number of shots required to achieve the baseline ST accuracy, while $\mathrm{SRF}<1$ indicates no net savings under this shot-noise-dominated conversion.  Values of $\mathrm{SRF}>N/m$ (marked by the dashed line) indicate when the net reduction surpasses the explicit reduction offered by time-sub-sampling, meaning that there is still room in the error budget which could be exchanged for further shot reduction relative to the baseline scheme.}
    \label{fig:rf}
\end{figure}

We translate the observed reduction in reconstruction error into an effective reduction in measurement shots by using the standard shot-noise scaling of Pauli-shadow estimators, $\mathrm{RMSE}\propto 1/\sqrt{\# \text{shots}}$.  Baseline ST uses $N_{\mathrm{ST}}$ snapshots \emph{per timestep} over $N$ timesteps, hence
$N_{\mathrm{tot}}^{\mathrm{ST}} = N N_{\mathrm{ST}}$.  For CSST, we sample $m$ timesteps and (in the reported sweeps) use the same per-timestep snapshot budget
$N_{\mathrm{ST}}$, optimizing the regularization parameter $\alpha$ in \cref{eq:lasso}.

For each observable $O_i$ (within a fixed Pauli-weight sector), define
\beq
    R_i \equiv \frac{\mathrm{RMSE}_i^{\mathrm{CS}}}{\mathrm{RMSE}_i^{\mathrm{ST}}},
\eeq
where $\mathrm{RMSE}_i^{\mathrm{ST}}$ is the baseline ST RMSE and $\mathrm{RMSE}_i^{\mathrm{CS}}$ is the best CSST RMSE (at the same per-timestep snapshot budget $N_{\mathrm{ST}}$, but using only $m$ sampled timesteps).  Assuming shot noise dominates (and other error sources are subleading), we model the RMSE at $\nu$ snapshots per timestep as
$\mathrm{RMSE}_i^{\mathrm{ST}}(\nu) \approx C_{\mathrm{ST}}/\sqrt{\nu}$ and 
$\mathrm{RMSE}_i^{\mathrm{CS}}(\nu) \approx C_{\mathrm{CS}}/\sqrt{\nu}$,
where $C_{\mathrm{ST}}$ and $C_{\mathrm{CS}}$ are (observable-dependent) prefactors. Since $R_i$ is measured at the same
$\nu=N_{\mathrm{ST}}$, it estimates the prefactor ratio:
\beq
    R_i  =  \frac{\mathrm{RMSE}_i^{\mathrm{CS}}(N_{\mathrm{ST}})}{\mathrm{RMSE}_i^{\mathrm{ST}}(N_{\mathrm{ST}})}
     \approx  \frac{C_{\mathrm{CS}}}{C_{\mathrm{ST}}}.
\eeq
We then ask how many snapshots per sampled timestep CSST would need to \emph{match the baseline ST RMSE}: define $N_{\mathrm{eff}}^{\mathrm{ST}}$ by
$\mathrm{RMSE}_i^{\mathrm{CS}} \left(N_{\mathrm{eff}}^{\mathrm{ST}}\right)
=
\mathrm{RMSE}_i^{\mathrm{ST}} \left(N_{\mathrm{ST}}\right)$.
Plugging this in the shot-noise models above gives
\beq
    \frac{C_{\mathrm{CS}}}{\sqrt{N_{\mathrm{eff}}^{\mathrm{ST}}}}
    =
    \frac{C_{\mathrm{ST}}}{\sqrt{N_{\mathrm{ST}}}}
    \quad\Rightarrow\quad
    N_{\mathrm{eff}}^{\mathrm{ST}} \approx \left(\frac{C_{\mathrm{CS}}}{C_{\mathrm{ST}}}\right)^2 N_{\mathrm{ST}}
    \approx R_i^2 N_{\mathrm{ST}}.
\eeq
Thus, when $R_i<1$, CSST would require only about $N_{\mathrm{eff}}^{\mathrm{ST}}\approx R_i^2 N_{\mathrm{ST}}$ snapshots per sampled timestep to match the baseline ST error.

With $N_{\mathrm{tot}}^{\mathrm{CS}} \equiv m N_{\mathrm{eff}}^{\mathrm{ST}}$, this motivates the inferred shot-reduction factor
\begin{align}
    \label{eq:SRF}
    \mathrm{SRF}_i
    \equiv \frac{N_{\mathrm{tot}}^{\mathrm{ST}}}{N_{\mathrm{tot}}^{\mathrm{CS}}}
    \approx
    \frac{N N_{\mathrm{ST}}}{m R_i^2 N_{\mathrm{ST}}}
    =
    \frac{N}{m}\cdot\frac{1}{R_i^2}.
\end{align}
Values $\mathrm{SRF}_i>1$ indicate a net shot savings relative to baseline ST at the same target RMSE, while $\mathrm{SRF}_i<1$ indicates no net savings. Since this conversion assumes RMSE is dominated by $1/\sqrt{\text{shots}}$ noise and neglects systematic errors (e.g., bias/model-mismatch or SPAM), we interpret $\mathrm{SRF}_i$ as an empirical \emph{effective} shot-reduction metric, complementary to the rigorous guarantee offered by \cref{eq:rf_theory}.  

One can interpret the factor $N/m$ as an explicit reduction in shots due to time-subsampling (measuring only $m$ of the $N$ timesteps at a fixed per-timestep snapshot budget). The factor $1/R_i^2$ can be interpreted as an implicit, counterfactual per-timestep shot reduction enabled by the sparsity-based reconstruction: under the shot-noise scaling $\mathrm{RMSE}\propto 1/\sqrt{\#\mathrm{shots}}$, an observed RMSE ratio $R_i$ at $N_{\mathrm{ST}}$ shots implies that CSST would require only $N_{\mathrm{eff}}^{\mathrm{ST}} = R_i^2 N_{\mathrm{ST}}$ snapshots per sampled timestep to reach the baseline ST RMSE. At $m=N$ in particular, $\mathrm{SRF}_i=1/R_i^2$ should therefore be read as a potential per-timestep shot reduction (i.e., a denoising gain), not as a literal reduction in the number of shots used in our numerical experiments.

We report weight-sector averages of $\mathrm{SRF}_i$ in \cref{fig:rf}.
As shown in this figure, the inferred shot-reduction factor exhibits a clear tradeoff between (i) the explicit time-subsampling advantage ($N/m$) and (ii) the reconstruction/denoising advantage captured by the error ratio $R_i$. For fixed per-timestep snapshot budget $N_{\mathrm{ST}}$, $\mathrm{SRF}$ generally decreases as $m$ increases, consistent with the fact that measuring more timesteps reduces the prefactor $N/m$ in \cref{eq:rf_theory}.  

However, importantly, the curves remain above $\mathrm{SRF}=1$ throughout the range of $m$ shown and are often above the reference scaling $\mathrm{SRF}=N/m$ (which corresponds to $R_i=1$).  Thus, the observed shot savings are not solely a consequence of time-subsampling: in many cases the sparse reconstruction also yields $R_i<1$, i.e., a lower RMSE than the baseline ST estimator at the same per-timestep snapshot budget.

At large $m$ (i.e., $m\simeq N$), the explicit subsampling factor is negligible and any advantage is therefore driven primarily by this denoising effect, $\mathrm{SRF}_i \approx 1/R_i^2$.  As $m$ decreases, some observables remain sufficiently sparse/compressible so that $R_i<1$ persists, whereas for others sparse recovery degrades and may yield $R_i\gtrsim 1$ (points above the diagonal in \cref{fig:cserrs_tind}).  Even in the latter case, a net shot saving can still occur because the explicit $N/m$ factor can offset a moderate increase in $R_i$: indeed,
\beq
    \mathrm{SRF}_i>1 \quad\Longleftrightarrow\quad R_i<\sqrt{N/m}.
\eeq
Equivalently, if one were to allocate the same total number of snapshots $m N_{\mathrm{ST}}$ to the baseline protocol (spread uniformly across all $N$ timesteps), the baseline would use only $(m/N)N_{\mathrm{ST}}$ snapshots per timestep and would therefore incur a larger shot-noise RMSE by a factor $\sqrt{N/m}$.

The dependence on $N_{\mathrm{ST}}$ is also pronounced (though note that for $w=1$ the dependence is non-monotonic). At small $N_{\mathrm{ST}}$ (shot-noise-dominated estimates), CSST yields the largest relative RMSE reductions (smaller $R_i$), and hence the largest inferred $\mathrm{SRF}$; these curves sit highest in \cref{fig:rf} and can remain well above unity even at large $m$. As $N_{\mathrm{ST}}$ increases, the baseline ST estimator becomes more accurate and there is less noise to be removed by sparse recovery, so $R_i\to 1$ and the inferred $\mathrm{SRF}$ decreases toward the no-denoising expectation (approaching $\mathrm{SRF}\to 1$ when $m=N$). 

Overall, \cref{fig:rf} indicates that CSST provides its strongest net shot savings in the low-to-moderate $N_{\mathrm{ST}}$ regime where shot noise is the dominant error source, while at high $N_{\mathrm{ST}}$ the benefit persists mainly as a smaller denoising gain.

\section{Conclusion}
\label{sec:conc}

Time-resolved measurement of many local observables is a recurring experimental bottleneck in quantum simulation and device characterization. In this work we addressed the task of estimating the full matrix of Pauli signals
$S_{ij}=\Tr \big(O_i\rho(t_j)\big)$
for a large family of low-weight Pauli strings $\{O_i\}_{i=1}^M$ across $N$ timesteps, with the goal of reducing the total number of device shots.

Our main contribution is a simple, modular protocol---\emph{Compressed Sensing Shadow Tomography} (CSST)---that combines two orthogonal ideas. Local classical shadows reduce the \emph{observable} dimension by allowing many Pauli expectation values to be estimated from the same randomized snapshot data at a fixed time. Compressed sensing then reduces the \emph{time} dimension by leveraging spectral structure in dynamics: many local expectation-value traces are well modeled (or well approximated) by superpositions of relatively few oscillatory and weakly decaying modes, and are therefore sparse or compressible in a Fourier- or cosine-type basis. 
Operationally, CSST samples a subset of $m\ll N$ timesteps uniformly at random (without replacement), collects local classical-shadow snapshots only at those times, forms estimates on the sampled grid, and reconstructs each length-$N$ signal via standard $\ell_1$-based recovery in the chosen unitary transform domain.  Experimentally, this involves a simple procedure of independent preparations and measurements at selected times, while shifting the burden to classical post-processing that can be implemented with existing convex-optimization tools.

On the theory side, we provide end-to-end guarantees by explicitly combining (i) finite-sample shadow estimation error at the sampled timesteps with (ii) standard compressed sensing stability bounds for time-subsampled signals. In the exactly sparse case, when each signal is $s$-sparse in a unitary transform basis (DFT/DCT), we show that
$m=\mathcal{O} \left(s\log^2 s \log N\right)$
random timesteps suffice (with high probability) for stable recovery, and we quantify how the required per-timestep snapshot count scales with the maximum Pauli weight, a target reconstruction accuracy (e.g., an $\mathrm{RMSE}$ constraint), and a uniform failure probability over $M$ observables. Compared to a baseline that runs shadow tomography independently at all $N$ timesteps with shots allocated uniformly across time, the resulting shot savings scale as $\widetilde{\Theta}(N/s)$, yielding a significant reduction whenever the effective sparsity level is much smaller than the time-series length. In the approximately sparse (compressible) setting, the reconstruction error naturally separates into a model-mismatch (tail) term and a noise term proportional to the effective measurement noise level, clarifying the tradeoff between sampling more times (improving conditioning and stability) versus taking more snapshots per sampled time (lowering the noise floor).

Our numerical experiments support the premise that physically generated Pauli signals can be strongly compressible in cosine/Fourier domains, and that accurate reconstructions are possible from a relatively small random fraction of timesteps. In particular, time-subsampling can achieve substantial shot reductions while maintaining (or even improving) reconstruction accuracy due to the denoising behavior of sparse recovery.

Several directions can strengthen and extend this framework. First, one can move beyond fixed Fourier/DCT bases to adaptive or model-based representations (e.g., windowed/multiresolution transforms, Prony/ESPRIT-style exponential models, or other physics-informed dictionaries). Second, recovering many observables independently leaves room for unexploited structure; taking advantage of shared spectral support (group sparsity), joint-sparsity/multiple-measurement-vector structure, or low-rank structure in the full signal matrix could further reduce both experimental and classical cost. Third, incorporating experimentally relevant effects such as drift across the acquisition window, SPAM errors, and correlated noise would sharpen the practical applicability of the guarantees. Finally, CSST is compatible with improved per-timestep measurement primitives (e.g., commuting-group measurements, derandomized or biased shadow ensembles), suggesting a route to compounded resource savings.

Overall, combining shadow tomography with compressed sensing provides a flexible route to reducing the scaling of time-resolved measurements. By explicitly leveraging spectral structure in dynamics while retaining the ``measure first, ask later'' advantages of shadows, CSST offers a systematic pathway toward high-resolution characterization of many observables under constrained shot budgets.

\section{Data and code availability}
The code \cite{barretoUSCqserverCSSTV1012026} and data \cite{barretoDataRepositoryCSST2026} of this work are openly available.

\section{Acknowledgements}
This material is based upon work supported by, or in part by, the U. S. Army Research Laboratory and the U. S. Army Research Office under contract/grant number W911NF2310255.  We thank Arman Babakhani and Onkar Apte for their feedback on early versions of this work.  

\bibliographystyle{apsrev4-2} 
\bibliography{CSST_paper} 

\newpage 
\clearpage

\appendix

\section{Notation}

\cref{tab:notation} provides a glossary of most symbols used in this work.

\begin{table*}[b]
\centering
\renewcommand{\arraystretch}{.91}
\begin{tabular}{|c|p{0.52\textwidth}|p{0.34\textwidth}|}
\hline
\textbf{Symbol} & \textbf{Meaning} & \textbf{Notes / dimensions} \\
\hline

\multicolumn{3}{|c|}{\textbf{Linear algebra and indexing}}\\
\hline
$a$ & scalar & \\
$\mathbf{a}$ & vector of scalars & typically $\mathbf{a}\in\mathbb{R}^N$ \\
$A$ & matrix of scalars & typically $A\in\mathbb{R}^{M\times N}$ \\
$\mathbf{A}$ & vector of matrices & e.g., $\mathbf{A}=(A_1,\dots,A_K)$ \\
$a_i=[\mathbf{a}]_i$ & element of vector $\mathbf{a}$ & \\
$\mathbf{a}_i=[A]_i$ & $i$th row of matrix $A$ & row vector in $\mathbb{R}^N$ \\
$A_i=[\mathbf{A}]_i$ & $i$th matrix in the vector $\mathbf{A}$ & \\
\hline

\multicolumn{3}{|c|}{\textbf{Quantum system, observables, and time series}}\\
\hline
$n$ & number of qubits & $d=2^n$ \\
$d$ & Hilbert-space dimension & $d=2^n$ \\
$\rho(t)$ & system state at time $t$ & density operator on $(\mathbb{C}^2)^{\otimes n}$ \\
$\rho_0$ & initial state & typically $\rho(0)$ \\
$O_i$ & $i$th target Pauli observable (Pauli string) & $i\in\{1,\dots,M\}$ \\
$w(O)$ & Pauli weight of observable $O$ & \# of non-identity single-qubit Paulis \\
$w$ & maximum Pauli weight in the target family & $w(O_i)\le w$ \\
$M$ & number of target observables & \\
$t_j$ & $j$th timestep & $j\in\{1,\dots,N\}$ \\
$N$ & number of timesteps in the full grid & \\
$S$ & Pauli-signal data matrix & $S\in\mathbb{R}^{M\times N}$ \\
$S_{ij}$ & expectation value entry & $S_{ij}\equiv \Tr \big(O_i\rho(t_j)\big)$ \\
$\mathbf{s}_i$ & length-$N$ signal for $O_i$ & $\mathbf{s}_i=(S_{i1},\dots,S_{iN})^T\in\mathbb{R}^N$ \\
$\widehat{\mathbf{s}}_i$ & estimate/reconstruction of $\mathbf{s}_i$ & from ST-at-all-times or CSST \\
$\mathrm{RMSE}(\mathbf{s}_i,\widehat{\mathbf{s}}_i)$ & RMSE over time & $\frac{1}{\sqrt{N}}\|\mathbf{s}_i-\widehat{\mathbf{s}}_i\|_2$ \\
$\epsilon_{\mathrm{RMS}}$ & target RMSE accuracy & $\mathrm{RMSE}\le\epsilon_{\mathrm{RMS}}\iff \|\mathbf{s}_i-\widehat{\mathbf{s}}_i\|_2\le \sqrt{N} \epsilon_{\mathrm{RMS}}$ \\
$\epsilon$ & target $\ell_2$ reconstruction accuracy & often $\epsilon=\sqrt{N} \epsilon_{\mathrm{RMS}}$ \\
$\delta$ & overall failure probability & uniform over observables (and time, as specified) \\
\hline

\multicolumn{3}{|c|}{\textbf{Time subsampling and compressed sensing}}\\
\hline
$m$ & number of sampled timesteps & $m\ll N$ \\
$\Omega$ & index set of sampled timesteps & $\Omega\subseteq\{1,\dots,N\}$, $|\Omega|=m$ \\
$F$ & unitary transform (DFT/DCT) & transform domain used for sparsity/compressibility \\
$s$ & sparsity level in transform domain & $F\mathbf{s}_i$ has $\le s$ nonzeros (or is compressible) \\
$P_\Omega$ & row-selection operator on indices $\Omega$ & $P_\Omega\in\{0,1\}^{m\times N}$ \\
$R_\Omega$ & normalized sampling operator & $R_\Omega=\sqrt{N/m} P_\Omega$ \\
$A$ & CS measurement matrix & $A=R_\Omega F^\dagger$ \\
$y$ & (rescaled) time-subsampled measurements & $y=R_\Omega \mathbf{s}+e$ \\
$e$ & additive noise in sampled measurements & induced by finite-shot ST estimation error \\
\hline

\multicolumn{3}{|c|}{\textbf{Shot complexity (measurement costs)}}\\
\hline
$N_{\mathrm{ST}}$ & shadow snapshots (shots) per measured timestep & “shot” and “snapshot” used interchangeably \\
$N_{\mathrm{tot}}^{\mathrm{ST}}$ & total shots for baseline ST-at-all-times & measures at all $N$ timesteps \\
$N_{\mathrm{tot}}^{\mathrm{CS}}$ & total shots for CSST (this work) & measures at only $m$ timesteps \\
\hline
\end{tabular}
\caption{Table of notation.}
\label{tab:notation}
\end{table*}

\newpage 
\clearpage

\section{Shadow estimators and error bounds}
\label{app:shadow}

A \emph{snapshot} is produced from a single measurement shot as follows:
\begin{enumerate}
    \item Select a random unitary $U$ from an ensemble $\mathcal{U}$.
    \item Apply $U$ to $\rho$.
    \item Measure each qubit in the computational ($Z$) basis to obtain a bitstring $b$, i.e., apply the projector $\ketbra{b}$.
    \item Form the snapshot
    \begin{align}
        \sigma \equiv U^\dagger \ketbra{b} U .
    \end{align}
\end{enumerate}

Averaging over both $U\sim\mathcal{U}$ and the Born-rule randomness in $b$ yields a linear measurement channel $\mathcal{M}$ such that
\begin{align}
    \mathbb{E}[\sigma]  =  \mathcal{M}(\rho).
\end{align}
If $\mathcal{M}$ is invertible, then applying $\mathcal{M}^{-1}$ to each snapshot defines a (classical) \emph{shadow} estimator
\begin{align}
    \hat{\rho}  \equiv  \mathcal{M}^{-1}(\sigma)
    = \mathcal{M}^{-1} \left(U^\dagger \ketbra{b} U\right),
    \quad
    \mathbb{E}[\hat{\rho}] = \rho.
\end{align}
This immediately yields an unbiased estimator for any observable $O$:
\begin{align}
    \hat{o}  \equiv  \Tr(O\hat{\rho}),
    \quad
    \mathbb{E}[\hat{o}] = \Tr(O\rho).
\end{align}

Everything after measurement must be efficiently computable classically: evaluating $\mathcal{M}^{-1}$, computing $\hat{o}$, and bounding $\mathbb{V}[\hat{o}]$ (or tail probabilities). The choice of ensemble $\mathcal{U}$ is what makes this efficient; for instance, Pauli shadows use a local measurement ensemble, while more structured schemes can leverage tensor networks~\cite{bertoniShallowShadowsExpectation2024, huDemonstrationRobustEfficient2025}.

\subsection{Pauli shadows}
For Pauli shadows, we choose
\begin{align}
    U = \bigotimes_{i=1}^n U_i,
    \quad
    U_i \in \{I,  H,  HS^\dagger\}
\end{align}
uniformly at random, and then measure in the $Z$ basis. (Equivalently, this corresponds to measuring a uniformly random full-weight Pauli string in $\{X,Y,Z\}^{\otimes n}$.)

In this case the measurement channel acts as a tensor product of single-qubit depolarizing channels with parameter $p=1/3$,
\begin{align}
    \mathcal{M}(X) = \bigl(\mathcal{D}_{1/3}\bigr)^{\otimes n}(X),
\end{align}
where for a single qubit
\begin{align}
    \mathcal{D}_{p}(X) = pX + (1-p)\frac{\Tr(X)}{2}I .
\end{align}
Its inverse is
\begin{align}
    \mathcal{D}_{1/3}^{-1}(X) = 3X - \Tr(X)I,
\end{align}
so the single-shot state estimator becomes
\begin{align}
    \hat{\rho}
    = \bigotimes_{j=1}^n \Bigl( 3  U_j^\dagger \ketbra{b_j}  U_j - I \Bigr).
\end{align}

If $O$ is a Pauli string, $O=\bigotimes_{j=1}^n P_j$ with $P_j\in\{I,X,Y,Z\}$, then
\begin{align}
    \hat{o} = \Tr(O\hat{\rho})
    = \prod_{j=1}^n \Bigl( 3\bra{b_j}U_j P_j U_j^\dagger \ket{b_j} - \Tr(P_j) \Bigr)
    = \prod_{j=1}^n W_j,
\end{align}
where
\begin{align}
    W_j =
    \begin{cases}
        1, & P_j = I,\\
        \pm 3, & U_j P_j U_j^\dagger = Z,\\
        0, & \text{otherwise.}
    \end{cases}
\end{align}
Thus, if $O$ has Pauli weight $w$, then $\hat{o}$ is nonzero only if the random choice of local bases diagonalizes $O$, which occurs with probability $3^{-w}$. In that event $\hat{o}=\pm 3^w$, so
\begin{align}
    \label{eq:EtoV}
    \mathbb{E}[\hat{o}^2] = 3^w
    \quad\Rightarrow\quad
    \mathbb{V}[\hat{o}] \le 3^w .
\end{align}

\subsection{Sample complexity}

Given $N_{\mathrm{snap}}$ i.i.d random variables $X_1,\dots,X_{N_{\mathrm{snap}}}$ where $a \leq X_i \leq b$ and $\hat{\mu}=\frac{1}{N_{\mathrm{snap}}}\sum_{i=1}^{N_{\mathrm{snap}}} X_i$, then Bernstein's inequality states
\begin{align}
    \label{eq:bern}
    \mathbb{P}[\left| \hat{\mu} - \mu \right| \geq \epsilon ] &\leq 2 \exp\left(-\frac{N_{\mathrm{snap}} \epsilon^2}{2\sigma^2 + \frac{1}{3} \Delta \epsilon} \right) \leq \delta \\
    \label{eq:bern2}
    \therefore N_{\mathrm{snap}} &\geq \frac{\log \left(2 /\delta \right)}{\epsilon^2}  \left( 2 \sigma^2 + \frac{1}{3} \Delta \epsilon \right)
\end{align}
where $\Delta \equiv b-a$.  The median-of-means estimator, formed by taking the median of $K$ $N_{\mathrm{snap}}/K$-sample means, obeys
\begin{align}
    \mathbb{P}\left[  |\hat{\mu}_{\text{MoM}} -  \mu| \geq \epsilon \right] &\leq \exp\left( - \frac{N_{\mathrm{snap}}\epsilon^2}{32 \sigma^2} \right) \leq \delta \\
    \label{eq:N_snap}
    \therefore N_{\mathrm{snap}} &\geq \frac{32 \log(1 /\delta)}{\epsilon^2} \sigma^2
\end{align}
provided $K=8\log(1/\delta)$.  

Now, let $X_1,\dots,X_{N_{\mathrm{snap}}}$ be i.i.d. copies of $\hat{o}$. 
Then Bernstein's inequality yields a sample complexity scaling as
\begin{align}
    \label{eq:N-from-Bernstein}
    N_{\mathrm{snap}} = \mathcal{O} \left(\frac{3^w}{\epsilon^2}\log\frac{1}{\delta}\right)
\end{align}
for estimating a fixed Pauli observable to accuracy $\epsilon$ with failure probability at most $\delta$; this is \cref{eq:N_ST_0} after applying a union bound (replacing $\delta$ by $\delta/M$) to obtain uniform accuracy over a family of $M$ observables.

To prove this result, take \cref{eq:bern} and apply it to the i.i.d. single-shot estimators
\beq
    X_i \equiv \hat{o}^{(i)},\quad \hat{\mu}=\frac{1}{N_{\mathrm{snap}}}\sum_{i=1}^{N_{\mathrm{snap}}} X_i,\quad \mu=\mathbb{E}[X_i]=\Tr(O\rho).
\eeq

From \cref{eq:EtoV}, we have $\mathbb{V}[\hat{o}] \le 3^w$,
so we may take $\sigma^2=\mathbb{V}[X_i]\le 3^w$.
Also, since $|X_i|\le 3^w$, we can choose $a=-3^w$, $b=3^w$, hence
\beq
    \Delta=b-a = 2\cdot 3^w.
\eeq

Since $\sigma^2\le 3^w$ and $\Delta = 2\cdot 3^w$, we have
\begin{align}
    \frac{\log(2/\delta)}{\epsilon^2}\left(2\sigma^2+\frac{1}{3}\Delta \epsilon\right)
    &\le \frac{\log(2/\delta)}{\epsilon^2}\left(2\cdot 3^w+\frac{1}{3}\cdot (2\cdot 3^w)\epsilon\right)\notag \\
    &= \frac{3^w\log(2/\delta)}{\epsilon^2}\left(2+\frac{2}{3}\epsilon\right).
\end{align}
Therefore, it follows from \cref{eq:bern2} that it suffices to take
\beq
    \label{eq:Bern-simple}
    N_{\mathrm{snap}} \ge \frac{3^w\log(2/\delta)}{\epsilon^2}\left(2+\frac{2}{3}\epsilon\right).
\eeq
In the usual regime of interest $\epsilon=\mathcal{O}(1)$ (in particular $\epsilon\le 1$), the parenthetical factor is $\mathcal{O}(1)$, and $\log(2/\delta)=\Theta(\log(1/\delta))$. Therefore \cref{eq:N-from-Bernstein} holds.

Proceeding with Bernstein's inequality in the form \cref{eq:Bern-simple} and with the union bound applied so that $\delta\to\delta/M$, let us take as an illustration the desired failure probability $\delta=10^{-2}$ and per-observable estimation error $\epsilon=10^{-2}$.  The number of Pauli strings of max weight $w$ in an $n$-qubit system is
\begin{align}
    M = \sum_{k=1}^{w} {n \choose k} 3^k
\end{align}
With $n=6$ and $w=4$, so that $M=1908$, we find that at least $N_{\mathrm{snap}} \geq  \frac{81}{10^{-4}}\ln\left(\frac{2\cdot 1908}{10^{-2}}\right)\left(2+\tfrac{2}{3}10^{-2}\right)\approx 2 \times 10^7$ shadow snapshots are needed.  Assuming a measurement time of $1\mu s$ and $N\sim 10^3$ timesteps at which one takes shadow snapshots, this corresponds to about $6$ hours of measurements, over which timescale the hardware parameters themselves might drift, necessitating recalibration cycles and additional overhead.  
 
If instead we use the median-of-means bound, i.e., \cref{eq:N_snap} with $\delta\to\delta/M$ (union bound) and $\sigma^2<3^w$, we obtain $N_{\mathrm{snap}} \geq 32\ln(10^2\cdot 1908)/10^{-4} 3^4 \approx 3\times 10^8$ snapshots, again illustrating that such worst-case tail bounds can be extremely conservative (though this can be tweaked by modifying the number of batches $K$). 

In any case, one need not rely on such bounds for practical recommendations since they are potentially loose; instead, in our numerical analysis we use at most a modest $50 \times 10^3$ shadow snapshots per timestep.

\section{Compressed Sensing}
\label{app:cs}

Here we follow \cite{foucartMathematicalIntroductionCompressive2013} and summarize the compressed-sensing facts used in this work, specialized to our setting.
We consider an unknown coefficient vector $x\in\mathbb{C}^N$ and $m$ measurements $y\in\mathbb{C}^m$ obtained through a measurement matrix $A\in\mathbb{C}^{m\times N}$:
\begin{align}
    y = Ax .
\end{align}
Without additional structure, recovering an arbitrary $x\in\mathbb{C}^N$ typically requires $m\ge N$ measurements.  Compressed sensing leverages \emph{sparsity} or \emph{compressibility} to allow recovery from $m\ll N$ measurements.

\subsection{Sparse and compressible signals}

A vector $x$ is \emph{$s$-sparse} if it has at most $s$ nonzero entries, i.e.,
\begin{align}
    \|x\|_0 \equiv \bigl|\{j\in\{1,\dots,N\}:x_j\neq 0\}\bigr| \le s.
\end{align}
More generally, many signals are not exactly sparse but are well-approximated by sparse vectors.
Let $|x|_{(1)}\ge |x|_{(2)}\ge \cdots \ge |x|_{(N)}$ denote the entries of $x$ sorted by magnitude.
Following \cite{candesStableSignalRecovery2005}, we say that $x$ is \emph{compressible} if there exist constants $C_r>0$ and $r>1$ such that
\begin{align}
    |x|_{(k)} \le C_r k^{-r},\quad k=1,2,\dots,N.
\end{align}
For any $s\in\{1,\dots,N\}$, let $x_s$ denote a best $s$-term approximation obtained by retaining the $s$ largest-magnitude entries of $x$ and setting the rest to zero (not necessarily unique).

A standard consequence of power-law decay is an explicit tail bound:
\begin{align}
    \label{eq:l1_tail_powerlaw_app}
    \|x-x_s\|_1
    = \sum_{k>s} |x|_{(k)}
    \le \sum_{k>s} C_r k^{-r}
    \le \frac{C_r}{r-1}  s^{ 1-r},
\end{align}
and similarly $\|x-x_s\|_2 \le \frac{C_r}{\sqrt{2r-1}}  s^{ \frac12-r}$ (see, e.g., \cite{foucartMathematicalIntroductionCompressive2013}).

\subsection{Recovery via $\ell_1$ minimization}

If the support of an $s$-sparse vector were known, one could restrict to the corresponding $s$ columns of $A$ and solve a reduced linear system.
In practice the support is unknown.
The ideal recovery procedure is $\ell_0$ minimization,
\begin{align}
    \label{eq:l0_app}
    \min_\mathbf{z} \|\mathbf{z}\|_0\quad \text{s.t.}\quad A \mathbf{z} = \mathbf{y},
\end{align}
which can recover sparse vectors under suitable conditions on $A$ but is NP-hard in general.
A standard tractable alternative is $\ell_1$ minimization.

In the noiseless case, \emph{Basis Pursuit} (BP) solves
\begin{align}
    \label{eq:bp_app}
    \min_\mathbf{z} \|\mathbf{z}\|_1\quad \text{s.t.}\quad A \mathbf{z} = \mathbf{y}.
\end{align}
In our setting, measurements are noisy: we observe $\mathbf{y}=A\mathbf{x}+\mathbf{e}$ where $\|\mathbf{e}\|_2\le \eta$, and $\mathbf{x}$ may be only approximately sparse.
The robust constrained formulation used in the main text is
\emph{Quadratically Constrained Basis Pursuit} (QCBP),
\begin{align}
    \label{eq:qcbp_app}
    \min_\mathbf{z} \|\mathbf{z}\|_1\quad \text{s.t.}\quad \|A\mathbf{z}-\mathbf{y}\|_2\le \eta.
\end{align}
Closely related alternatives include the penalized and $\ell_1$-constrained LASSO-type variants [\cref{eq:lasso} in the main text],
\begin{align}
    \label{eq:bpd_app}
    \min_\mathbf{z} \ \frac{1}{2}\|A\mathbf{z}-\mathbf{y}\|_2^2 + \lambda \|\mathbf{z}\|_1,
    \quad
    \min_\mathbf{z} \ \frac{1}{2}\|A\mathbf{z}-\mathbf{y}\|_2^2 \ \ \text{s.t.}\ \ \|\mathbf{z}\|_1 \le \tau,
\end{align}
which are equivalent to \cref{eq:qcbp_app} in the usual Lagrange-multiplier sense (i.e., for appropriate choices of $\lambda$ or $\tau$ one recovers the same solution set as QCBP).
Naming conventions vary in the literature; to match the main text, we reserve QCBP for the constrained form \cref{eq:qcbp_app} and refer to \cref{eq:bpd_app} as the penalized/constrained BPDN/LASSO-type formulations.

\subsection{Restricted Isometry Property and stability guarantees}

A convenient sufficient condition for uniform recovery is the Restricted Isometry Property (RIP); see \cref{def:rip}.
In particular, if $A$ satisfies RIP of order $2s$ with constant $\delta_{2s}<1$, then $A$ is injective on $2s$-sparse vectors, so distinct $s$-sparse signals cannot map to the same measurement vector.

We use the same stability guarantee as in the main text (Theorem~\ref{th:cs}).
If $A$ satisfies RIP of order $2s$ with
\begin{align}
    \delta_{2s} < \frac{4}{\sqrt{41}},
\end{align}
and if $\mathbf{x}\in\mathbb{C}^N$ and $\mathbf{y}=A\mathbf{x}+\mathbf{e}$ with $\|\mathbf{e}\|_2\le \eta$, then any minimizer $\mathbf{x}^*$ of QCBP \cref{eq:qcbp_app} obeys
\begin{align}
    \label{eq:cs_stability_app}
    \|\mathbf{x}-\mathbf{x}^*\|_2 \le c_1 \frac{\|\mathbf{x}-\mathbf{x}_s\|_1}{\sqrt{s}} + c_2 \eta,
\end{align}
where $\mathbf{x}_s$ retains the $s$ largest (in magnitude) entries of $\mathbf{x}$ and $c_1,c_2>0$ depend only on the RIP constant (see \cite{candesStableSignalRecovery2005, candesIntroductionCompressiveSampling2008}).
Thus, the recovery error decomposes into a \emph{compressibility error} term (vanishing when $\mathbf{x}$ is exactly $s$-sparse) and a \emph{noise amplification} term proportional to the measurement noise level $\eta$.

\subsection{RIP for subsampled Fourier/DCT measurements}

In our application, a time-domain signal $\mathbf{s}\in\mathbb{C}^N$ is assumed sparse/compressible in an orthonormal transform
domain $F\in\mathbb{C}^{N\times N}$ (e.g., the DFT or an orthonormal DCT), i.e., $\mathbf{x}=F\mathbf{s}$.
If we sample $\mathbf{s}$ at a uniformly random subset $\Omega\subset\{1,\dots,N\}$ of size $m$,
then the (normalized) measurement matrix acting on the coefficient vector $\mathbf{x}$ is
\begin{align}
    A = \sqrt{\frac{N}{m}}  P_\Omega F^\dagger,
\end{align}
where $P_\Omega\in\{0,1\}^{m\times N}$ is the row-selection operator (so $P_\Omega \mathbf{v}$ keeps only entries of $\mathbf{v}$ indexed by $\Omega$).

The prefactor $\sqrt{N/m}$ is required to match the normalization used in standard RIP analyses.
To see this, note that for any $\mathbf{z}\in\mathbb{C}^N$,
\begin{align}
    \mathbb{E}_\Omega \bigl\|P_\Omega \mathbf{z}\bigr\|_2^2
    = \mathbb{E}_\Omega  \left[\sum_{j\in\Omega}|z_j|^2\right]
    = \frac{m}{N}\sum_{j=1}^N |z_j|^2
    = \frac{m}{N} \|\mathbf{z}\|_2^2.
\end{align}
In particular, if $\mathbf{z}=F^\dagger \mathbf{x}$ with $F$ unitary, then
$\mathbb{E}_\Omega\|P_\Omega F^\dagger \mathbf{x}\|_2^2=(m/N)\|\mathbf{x}\|_2^2$.
Thus, even in the ideal noiseless setting, the unnormalized operator $P_\Omega F^\dagger$
can at best act as an approximate isometry up to a global factor $m/N$.
The normalization $A=\sqrt{N/m} P_\Omega F^\dagger$ ensures $\mathbb{E}_\Omega\|A \mathbf{x}\|_2^2=\|\mathbf{x}\|_2^2$,
matching the hypotheses of standard RIP results for subsampled bounded-unitary matrices.

A standard guarantee for such subsampled bounded-unitary systems is the following:
if $V\in\mathbb{C}^{N\times N}$ is unitary and satisfies $\|V\|_\infty \le C/\sqrt{N}$, and $A$
is formed by sampling $m$ rows uniformly at random from $V$ and multiplying by $\sqrt{N/m}$,
then for
\begin{align}
    m = \mathcal{O}\left(\delta_k^{-4} k \log^2\Big(\frac{k}{\delta_k}\Big)\log N\right),
\end{align}
the matrix $A$ satisfies RIP of order $k$ with constant $\delta_k$ with high probability
\cite{havivRestrictedIsometryProperty2015}.
In our applications, $V=F^\dagger$ with $F$ equal to the DFT (for which $C=1$) or an orthonormal DCT-II (for which $C=\sqrt{2}$),
so the boundedness condition holds.

In the main text we apply this result with $k=2s$ and $\delta_{2s}=4/\sqrt{41}-\kappa$ (for any fixed $0<\kappa\ll 1$), yielding the sampling rate
\begin{align}
    \label{eq:m-again}
    m = \mathcal{O}\left(s\log^2 s\log N\right)
\end{align}
sufficient for stable compressed sensing recovery of $s$-sparse (or $s$-compressible) signals
from randomly time-subsampled measurements.

As in the main text, if one wants the RIP event to hold with probability at least $1-\delta_{\mathrm{RIP}}$, one may increase the implicit constant in \cref{eq:m-again} so that the failure probability in the high-probability bound is at most $\delta_{\mathrm{RIP}}$; we absorb this dependence into polylogarithmic factors.

\section{Asymptotic notation in shot-count comparisons}
\label{app:asymp_shots}

We use standard asymptotic notation: for nonnegative functions $f,g$ of problem parameters, $f=\mathcal{O}(g)$ means that there exists a constant $C>0$ such that $f \le C g$ for all parameter values of interest, while $f=\Omega(g)$ means $f \ge c g$ for some constant $c>0$, and $f=\Theta(g)$ means both $f=\mathcal{O}(g)$ and $f=\Omega(g)$. We also use $\widetilde{\mathcal{O}}(\cdot)$ and $\widetilde{\Theta}(\cdot)$ to suppress polylogarithmic factors in the relevant parameters.

In this paper, many shot-count expressions are stated as sufficient conditions derived from concentration bounds.
For example, statements of the form
\beq
    N_{\mathrm{tot}}^{\mathrm{ST}} = \mathcal{O}(f(\cdot)),\qquad
    N_{\mathrm{tot}}^{\mathrm{CS}} = \mathcal{O}(g(\cdot))
\eeq
should be read as: there exist absolute constants $C_{\mathrm{ST}},C_{\mathrm{CS}}>0$ (independent of $N,m,s,M,\delta$ and the target accuracies) such that taking
$N_{\mathrm{tot}}^{\mathrm{ST}} = C_{\mathrm{ST}} f(\cdot)$ and $N_{\mathrm{tot}}^{\mathrm{CS}} = C_{\mathrm{CS}} g(\cdot)$
is sufficient to achieve the stated accuracy and failure probability. These bounds do \emph{not} claim that the resulting shot budgets are information-theoretically optimal; proving optimality would require matching lower bounds (i.e., $\Omega(\cdot)$ statements).

When we compare two such sufficient budgets, the ratio is therefore controlled up to absolute constant factors:
\beq
    \frac{N_{\mathrm{tot}}^{\mathrm{ST}}}{N_{\mathrm{tot}}^{\mathrm{CS}}}
    =
    \frac{C_{\mathrm{ST}}}{C_{\mathrm{CS}}}\,\frac{f(\cdot)}{g(\cdot)}
    =
    \Theta\left(\frac{f(\cdot)}{g(\cdot)}\right),
\eeq
where the implicit constants in $\Theta(\cdot)$ are given by the constant ratio $C_{\mathrm{ST}}/C_{\mathrm{CS}}$.
Thus, uses of $\Theta(\cdot)$ [or $\widetilde{\Theta}(\cdot)$] in shot-reduction statements should be interpreted as comparisons of the prescribed sufficient shot budgets, up to absolute constants (and, for tilded notation, up to polylogarithmic factors), rather than as claims about the exact optimal sample complexity.

\section{Derivation of the tail bound in the approximately sparse case}
\label{app:approx-sparse}

Our goal in this appendix is to derive \cref{eq:s-s*}. As in the main text, consider a single real component sampled on $n=0,\dots,N-1$:
\begin{equation}
    s_n  =  e^{-\gamma n\Delta t}\cos(\omega n\Delta t+\phi).
    \label{eq:damped_cosine}
\end{equation}
Let $T=N\Delta t$ and let $F$ be the orthonormal DCT-II matrix,\footnote{The (type-II)
discrete cosine transform (DCT-II) is the orthonormal linear map $F\in\mathbb{R}^{N\times N}$
with entries
$
F_{kn}=\sqrt{\frac{2-\delta_{k0}}{N}}\cos \bigl(\frac{\pi}{N}(n+\frac12)k\bigr),
$
where $0\le n,k\le N-1$.}
and define the coefficient vector $x\equiv Fs$, i.e.,
\begin{equation}
    x_k  =  (Fs)_k  =  \sqrt{\frac{2-\delta_{k0}}{N}}\sum_{n=0}^{N-1}
    e^{-\gamma n\Delta t}\cos(\omega n\Delta t+\phi) 
    \cos \Bigl(\frac{\pi}{N}\Bigl(n+\frac12\Bigr)k\Bigr).
    \label{eq:dct_def}
\end{equation}
It is convenient to introduce the dimensionless parameters
\beq
    \Gamma\equiv \gamma\Delta t,
    \quad \alpha_k\equiv \frac{\pi k}{N},\quad \beta_k\equiv \frac{\pi k}{2N}, \quad z_{k,\pm}\equiv -\Gamma+i(\omega\Delta t\pm\alpha_k) .
\eeq
Then, using elementary trigonometric identities and summing the geometric series, we obtain
\begin{align}
    x_k
    &=\frac12\sqrt{\frac{2-\delta_{k0}}{N}} 
    \Re \left[
    e^{i(\phi-\beta_k)}\frac{1-e^{N z_{k,-}}}{1-e^{z_{k,-}}}
     + 
    e^{i(\phi+\beta_k)}\frac{1-e^{N z_{k,+}}}{1-e^{z_{k,+}}}
    \right].
    \label{eq:xk_geom}
\end{align}

From $|\Re(\cdot) |\le|\cdot|$ and the triangle inequality,
\begin{equation}
    |x_k|
     \le \frac12\sqrt{\frac{2-\delta_{k0}}{N}}\left(
    \left|\frac{1-e^{N z_{k,-}}}{1-e^{z_{k,-}}}\right|
    +
    \left|\frac{1-e^{N z_{k,+}}}{1-e^{z_{k,+}}}\right|
    \right).
    \label{eq:xk_bound_split}
\end{equation}
Moreover, since $\Re(z_{k,\pm})=-\Gamma\le 0$,
\begin{equation}
    |1-e^{N z_{k,\pm}}|\le 1+e^{N\Re(z_{k,\pm})}\le 2,
    \quad
    \sum_{n=0}^{N-1}|e^{n z_{k,\pm}}|\le \sum_{n=0}^{N-1}e^{-\Gamma n}\le N,
\end{equation}
so the geometric sums obey the uniform bound
\begin{equation}
    \left|\frac{1-e^{N z_{k,\pm}}}{1-e^{z_{k,\pm}}}\right|
    =\left|\sum_{n=0}^{N-1}e^{n z_{k,\pm}}\right|
    \le \min \left\{ N, \frac{2}{|1-e^{z_{k,\pm}}|}\right\}.
    \label{eq:geom_min_bound}
\end{equation}
Now note that, defining $\vartheta\equiv \omega\Delta t$ and using $1-\cos\theta=2\sin^2(\theta/2)$,
\begin{align}
    |1-e^{z_{k,\pm}}|^2
    &=\left|1-e^{-\Gamma}e^{i(\vartheta\pm\alpha_k)}\right|^2 \nonumber\\
    &=(1-e^{-\Gamma})^2+4e^{-\Gamma}\sin^2 \Bigl(\frac{\vartheta\pm\alpha_k}{2}\Bigr).
    \label{eq:denom_identity}
\end{align}
Combining Eqs.~\eqref{eq:xk_bound_split}-\eqref{eq:denom_identity} yields the bound
\begin{equation}
    |x_k|
     \le 
    \frac12\sqrt{\frac{2-\delta_{k0}}{N}}
    \sum_{\sigma\in\{-,+\}}
    \min \left\{ N, 
    \frac{2}{
    \sqrt{(1-e^{-\Gamma})^2+4e^{-\Gamma}\sin^2 \bigl(\frac{\vartheta+\sigma\alpha_k}{2}\bigr)}
    }\right\}.
    \label{eq:envelope_exact}
\end{equation}

Now consider the branch $\sigma=-$ and define the grid frequencies
\beq
    \nu_k\equiv \frac{\pi k}{T}\quad\text{so that}\quad \alpha_k=\nu_k\Delta t.
\eeq
Then $\vartheta-\alpha_k=(\omega-\nu_k)\Delta t$. When $\Gamma\ll 1$
(weak damping per timestep) and $|\vartheta-\alpha_k|\ll 1$ (near-resonant $k$),
we have the Taylor approximations
\beq
    1-e^{-\Gamma}=\Gamma+O(\Gamma^2),
    \quad
    \sin\Bigl(\frac{\vartheta-\alpha_k}{2}\Bigr)=\frac{\vartheta-\alpha_k}{2}+O\bigl((\vartheta-\alpha_k)^3\bigr),
\eeq
so the denominator in \cref{eq:envelope_exact} takes the Lorentzian-like form
\begin{equation}
    \sqrt{(1-e^{-\Gamma})^2+4e^{-\Gamma}\sin^2 \Bigl(\frac{\vartheta-\alpha_k}{2}\Bigr)}
     = 
    \sqrt{\Gamma^2+(\vartheta-\alpha_k)^2} \bigl(1+o(1)\bigr).
    \label{eq:lorentz_local}
\end{equation}
Equivalently, 
$\sqrt{\Gamma^2+(\vartheta-\alpha_k)^2}
=\Delta t \sqrt{\gamma^2+(\omega-\nu_k)^2}$, so the dominant DCT coefficients are
peaked around the closest grid frequency $\nu_\ell$ (defined below), with a width governed by
$\gamma$ (damping) and the finite observation window (through $N$).

The $\sigma=+$ branch corresponds to $\vartheta+\alpha_k$.  For $\vartheta\in(0,\pi)$ and $k\in\{0,\dots,N-1\}$ one has $\vartheta+\alpha_k\in(0,2\pi)$, so unless parameters conspire to place $\vartheta+\alpha_k$ extremely close to a multiple of $2\pi$, this term stays bounded and contributes only a smooth background compared to the sharp peak from $\sigma=-$. In what follows we focus on the $\sigma=-$ peak, which controls compressibility.

Let $\ell\in\{0,\dots,N-1\}$ be the closest grid index to the signal frequency, i.e.,
\begin{equation}
    \ell \in \arg\min_{0\le k\le N-1}|\vartheta-\alpha_k|
    \quad\Longleftrightarrow\quad
    \nu_\ell=\frac{\pi\ell}{T}\ \text{is the closest grid frequency to}\ \omega.
    \label{eq:closest_grid}
\end{equation}
For $k\neq \ell$, 
\begin{equation}
    |\vartheta-\alpha_k|
     \ge 
    \left||k-\ell|-\frac12\right|\frac{\pi}{N}
     \ge 
    \frac{\pi}{2N} |k-\ell|.
    \label{eq:mismatch_lower}
\end{equation}
Moreover, since $|\vartheta-\alpha_k|\le \pi$ for all $k$ and
$|\sin(u)|\ge \frac{2}{\pi}|u|$ for $|u|\le \frac{\pi}{2}$, we have
\begin{equation}
    \left|\sin \Bigl(\frac{\vartheta-\alpha_k}{2}\Bigr)\right|
     \ge \frac{1}{\pi}|\vartheta-\alpha_k|.
    \label{eq:sin_lower}
\end{equation}
Dropping the nonnegative $(1-e^{-\Gamma})^2$ term in \cref{eq:envelope_exact} and using
\cref{eq:sin_lower} gives, for the dominant $\sigma=-$ branch,
\begin{equation}
    \frac{2}{\sqrt{(1-e^{-\Gamma})^2+4e^{-\Gamma}\sin^2 \bigl(\frac{\vartheta-\alpha_k}{2}\bigr)}}
     \le 
    \frac{1}{\sqrt{e^{-\Gamma}} \left|\sin \bigl(\frac{\vartheta-\alpha_k}{2}\bigr)\right|}
     \le 
    \frac{\pi e^{\Gamma/2}}{|\vartheta-\alpha_k|}.
    \label{eq:denom_upper}
\end{equation}
Combining \cref{eq:envelope_exact}, \cref{eq:denom_upper}, and \cref{eq:mismatch_lower}
(and absorbing the $\sigma=+$ background into constants) yields the decay estimate
\begin{equation}
    |x_k|
     \le 
    C \sqrt{N} \frac{1}{|k-\ell|}\quad (k\neq \ell),
    \label{eq:harmonic_decay}
\end{equation}
for a constant $C$ that depends at most on $\Gamma$ through a harmless factor $e^{\Gamma/2}$
(and on the choice of treating the $\sigma=+$ term as a bounded background).

To convert \cref{eq:harmonic_decay} into a tail bound, we compare against an explicit
$s$-term truncation supported on a window around the peak index $\ell$.
Let
\begin{equation}
    r \equiv \left\lfloor \frac{s-1}{2}\right\rfloor,
    \quad
    I_s \equiv \{k:\ |k-\ell|\le r\}.
\end{equation}
Then $|I_s|\le 2r+1\le s$. Let $x_{I_s}$ denote the vector that agrees with $x$ on $I_s$
and is zero elsewhere. Using \cref{eq:harmonic_decay}, we obtain
\begin{align}
    \|x-x_{I_s}\|_1
    &=\sum_{k\notin I_s}|x_k|
     \le  2C\sqrt{N}\sum_{j=r+1}^{N-1}\frac{1}{j} \notag\\
    &\le 2C\sqrt{N}\left(\log\frac{N}{r+1}+1\right)
     \le 2C\sqrt{N}\left(\log\frac{2N}{s}+1\right)
     = \mathcal{O} \left(\sqrt{N}\log\frac{N}{s}\right),
    \label{eq:l1_tail_window}
\end{align}
where we used the standard harmonic-series bound and the fact that $r+1\ge s/2$.

Since $x_s$ is obtained by retaining the $s$ largest-magnitude entries of $x$, it minimizes
the $\ell_1$ tail among all truncations supported on at most $s$ indices. In particular,
\begin{equation}
    \|x-x_s\|_1 \le \|x-x_{I_s}\|_1.
\end{equation}
Therefore we obtain the claimed compressibility bound
\begin{equation}
    \|x-x_s\|_1
     = \|Fs-(Fs)_s\|_1
     \le  \mathcal{O} \left(\sqrt{N}\log\frac{N}{s}\right).
    \label{eq:l1_tail_bound}
\end{equation}
Substituting \cref{eq:l1_tail_bound} into \cref{eq:cs_error} finally yields
\begin{equation}
    \|s-s^\ast\|_2
     \le 
    O \left(\sqrt{\frac{N}{s}}\log\frac{N}{s}\right)
     + 
    O \left(\sqrt{N} \epsilon_{\mathrm{ST}}\right) ,
    \label{eq:damped_bound_final}
\end{equation}
which is \cref{eq:s-s*}.

\section{Additional data}
\label{app:additional}

\subsection{Reconstruction examples}
\label{app:examples}

\begin{figure}
    \centering
    \includegraphics[width=0.98\linewidth]{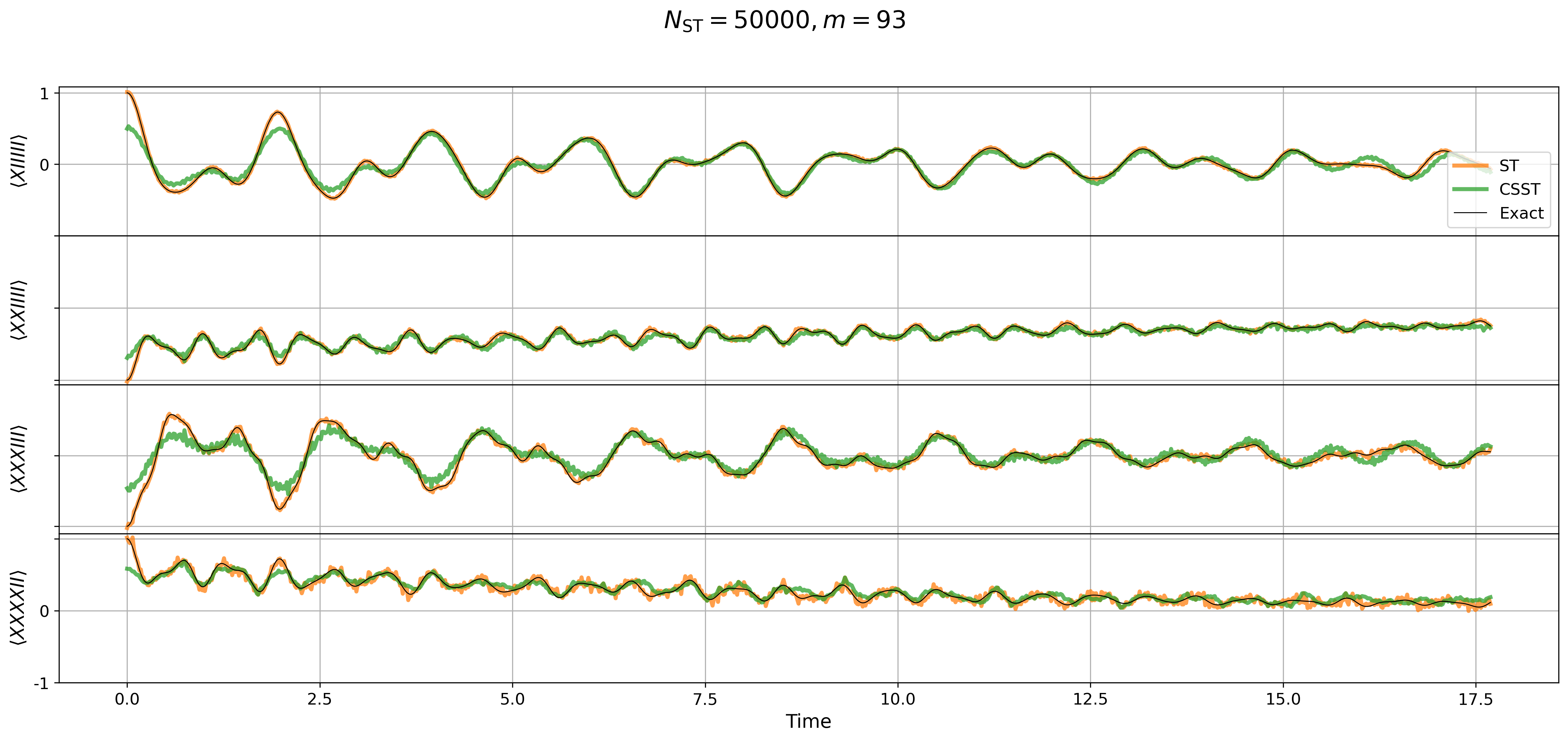}
    \includegraphics[width=0.98\linewidth]{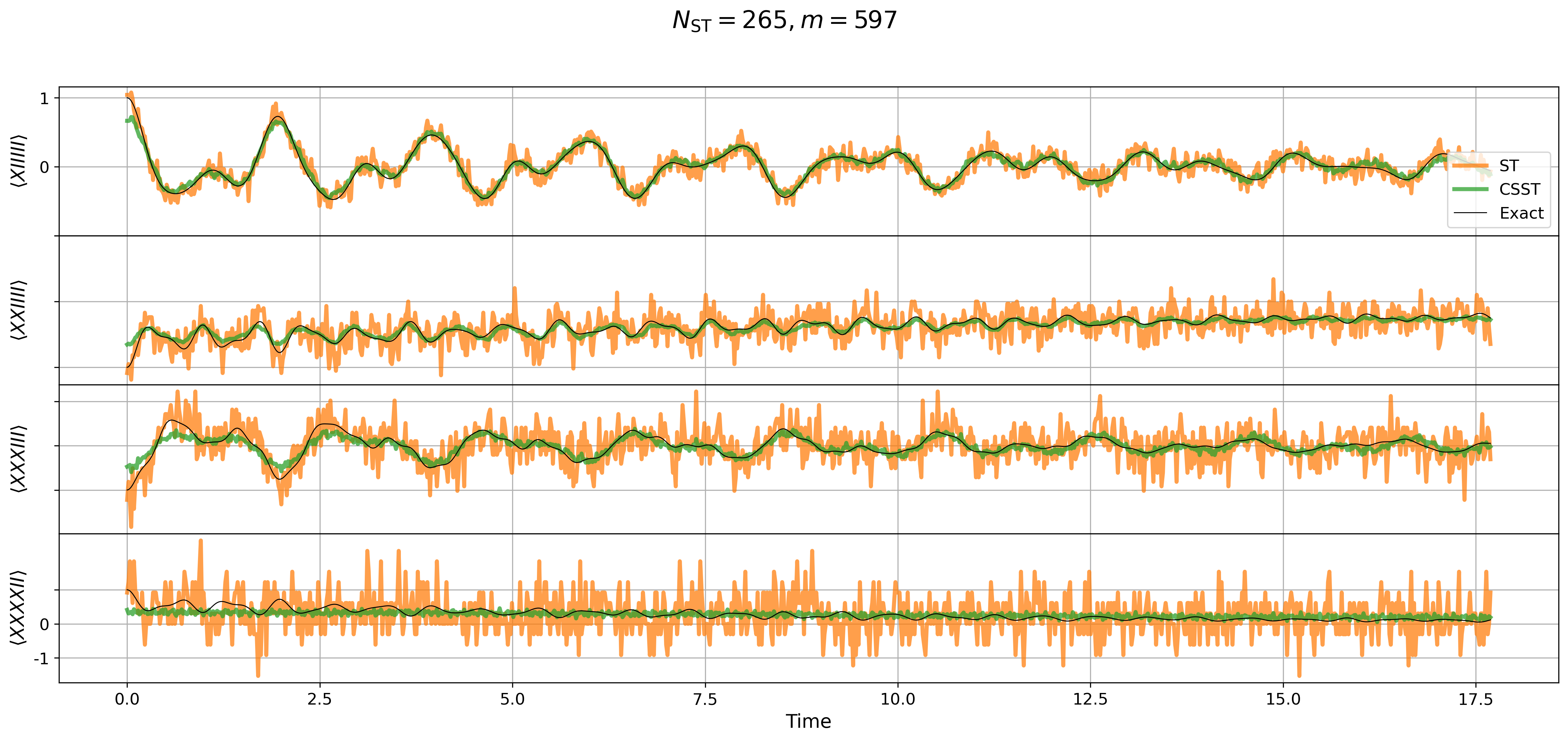}
    \caption{For two different parameter regimes, we plot $\langle O(t) \rangle$ over time as estimated via CSST at $\alpha^*$ from each observable's $\alpha$ sweep.  Top: we plot the data at $(N_{\mathrm{ST}}=50000, m=93)$ which represents one extreme where the shadow estimates are of high quality, so that even if only $\sim10$\% of the timesteps are retained, the signals are faithfully reconstructed.  Bottom: we plot the data at $(N_{\mathrm{ST}}=265, m=597)$ which represents the opposite extreme where few shadows are obtained per timesteps, and find that the dominant frequency and decay components are \textit{still} captured by the reconstruction when $\sim60$\% of the timesteps are retained (though at weight $4$, the SNR is around $-6$ and only the mean is recovered).}
    \label{fig:cs_examples}
\end{figure}

Here we provide additional CSST examples for selected observables in different parameter sweep regimes to complement \cref{fig:cs_example}.  Specifically, in \cref{fig:cs_examples}, we plot four observables as reconstructed by CSST ($\langle XIIIII \rangle$, $\langle XXIIII \rangle$, $\langle XXXIII \rangle$, $\langle XXXXII \rangle$) for the $2 \times 3$ Heisenberg model with initial state $|+-+-+-\rangle$, which is the running example throughout the main text.

\newpage
\subsection{Transverse-field Ising model}
\label{app:tfim}

\begin{figure*}
    \centering
    
    \makebox[\linewidth][c]{%
      \includegraphics[width=0.3\linewidth]{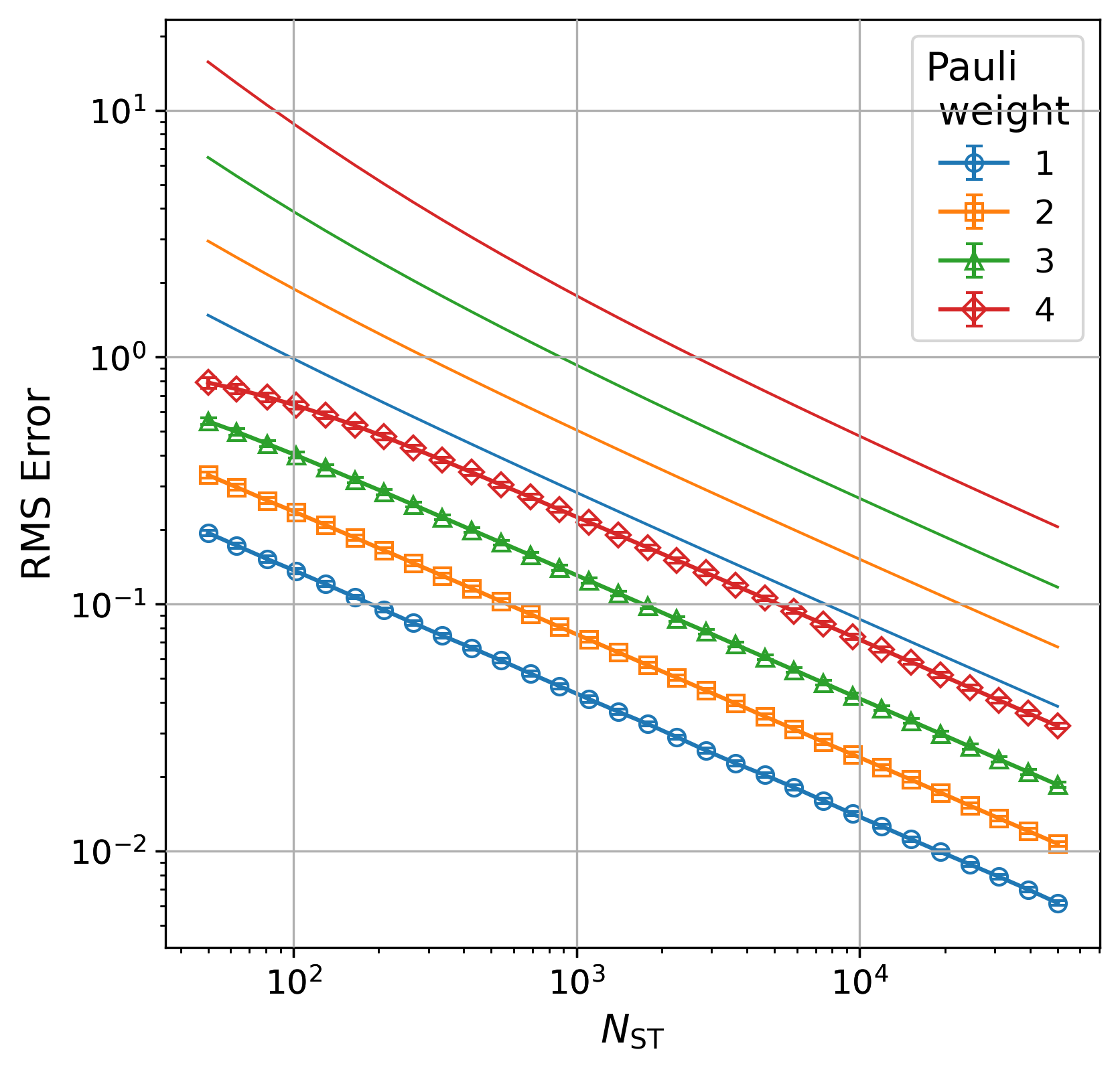}\hspace{0.05\linewidth}%
      \includegraphics[width=0.3\linewidth]{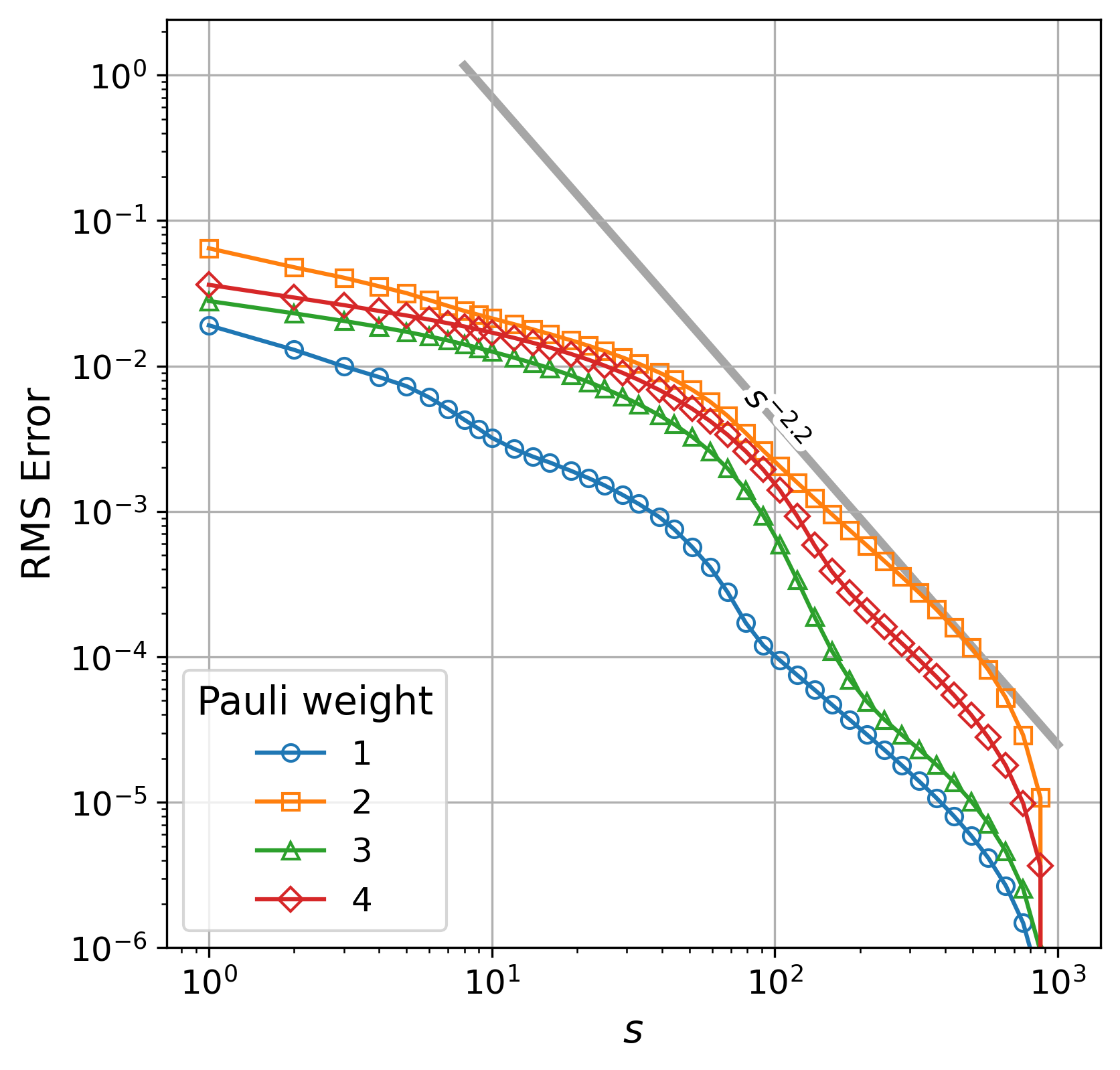}%
    }
    
    \par\medskip
    \includegraphics[width=0.9\linewidth]{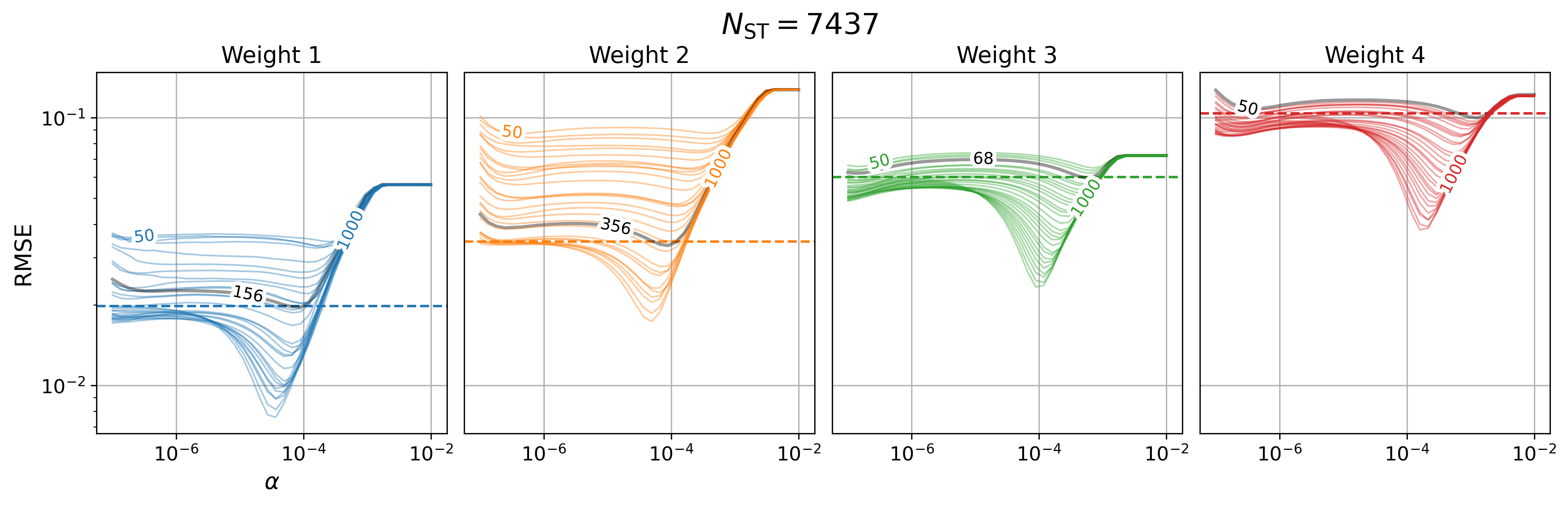}\par
    \includegraphics[width=0.9\linewidth]{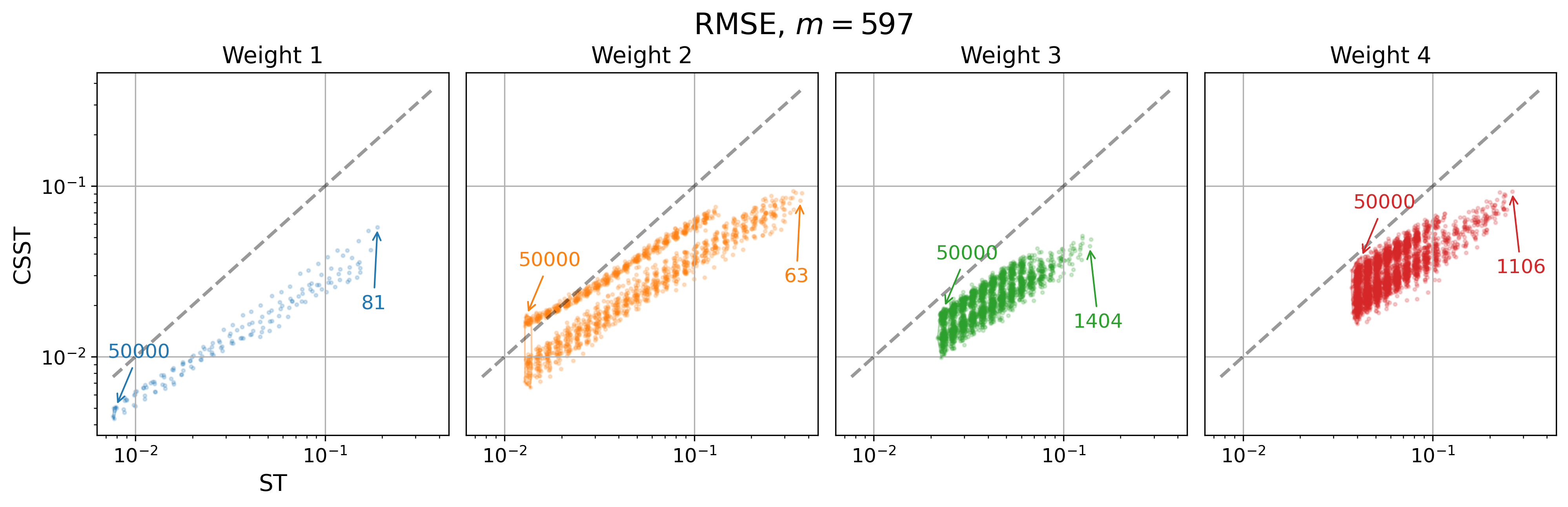}\par
    \includegraphics[width=0.9\linewidth]{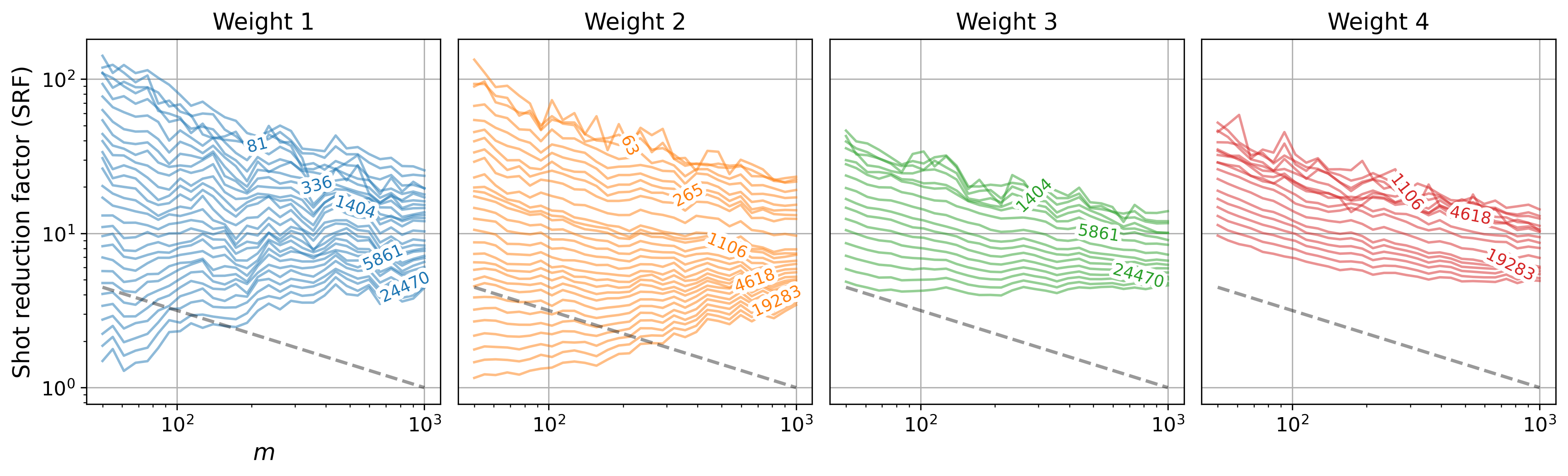}
    
    \caption{All simulations which appear in the main text are repeated here for the TFIM.  Different colors/data markers denote different Pauli weights in all plots.  For comparison, the top row corresponds to \cref{fig:setup}, upper middle to \cref{fig:cserrs_sind}, lower middle to \cref{fig:cserrs_tind}, and bottom to \cref{fig:rf}.  The $\alpha^*$ (optimal $\alpha$) behavior is already included in the main text, in the right panel of \cref{fig:optalphas}.  The qualitative behavior across all plots is comparable to that of the Heisenberg model in the main text, emphasizing that our conclusions are not limited to a specific choice of Hamiltonian or initial state.}
    \label{fig:2x3_tfim_ghz_app}
\end{figure*}

In this section, we provide additional data for a noisy transverse-field Ising model (TFIM) with initial state $|\text{GHZ}\rangle$ to complement the running example in the main text.  The model's Hamiltonian is
\begin{align}
    H = J \sum_{\langle a,b\rangle}Z_aZ_b + h \sum_a X_a,
\end{align}
where we set $J=h=1$, and the model has nearest-neighbor connectivity on a square lattice with open boundary conditions.  The initial state is
\begin{align}
    |\text{GHZ}\rangle = \frac{1}{\sqrt{6}} \left( |000000\rangle + |111111\rangle \right).
\end{align}
The dissipation is modeled just as in \cref{eq:lme}, i.e. with uniform single-qubit dephasing and amplitude damping rates $\gamma_\phi=\gamma_1=\gamma=10^{-1}$.  The plots in \cref{fig:2x3_tfim_ghz_app} are of the same type as those which appear in the main text, and have similar interpretations.

\end{document}